\documentclass[prb,twocolumn,superscriptaddress,aps]{revtex4-1}
\usepackage{color}
\usepackage{amsmath}
\usepackage{dcolumn}% Align table columns on decimal point
\usepackage{bm}% bold math
\usepackage{slashed}
\usepackage{textcomp}
\usepackage[toc,page]{appendix}
\usepackage{float}
\usepackage{amssymb}

\usepackage[pdftex]{graphicx}

\def\gapp{\lower.35em\hbox{$\stackrel{\textstyle>}{\sim}$}}
\def\lapp{\lower.35em\hbox{$\stackrel{\textstyle<}{\sim}$}}

\begin{document}
\bibliographystyle{apsrev4-1}
%

%\draft
\title{Symmetries and selection rules in the measurement of the phonon spectrum of graphene and
related materials}

%\title{Selection rules in the measurement of the phonon
%spectrum of graphene}

\author{Fernando de Juan}
\affiliation{Materials Science Division, Lawrence Berkeley National Laboratories, Berkeley, CA
94720, USA}
\affiliation{Department of Physics, University of California, Berkeley, CA 94720, USA}
\author{Antonio Politano}
\affiliation{Dipartimento di Fisica, Universit\`a degli Studi della Calabria, 87036 Rende
(Cs), Italy}
\author{Gennaro Chiarello}
\affiliation{Dipartimento di Fisica, Universit\`a degli Studi della Calabria, 87036 Rende
(Cs), Italy}
\affiliation{Consorzio Nazionale Interuniversitario di Scienze Fisiche della Materia, via della
Vasca Navale 84, 00146, Rome, Italy}
\author{Herbert A. Fertig}
\affiliation{Department of Physics, Indiana University, Bloomington, IN 47405, USA}

\date{\today}

\begin{abstract}
When the phonon spectrum of a material is measured in a scattering experiment, selection rules
preclude the observation of phonons that are odd under reflection by the scattering plane.
Understanding these rules is crucial to correctly interpret experiments and to
detect broken symmetries. Taking graphene as a case study, in this work we derive the complete set
of selection rules for the honeycomb lattice, showing that some of them have been missed or
misinterpreted in the literature. Focusing on the technique of high-resolution electron energy loss
spectroscopy (HREELS), we calculate the scattering intensity for a simple force constant
model to illustrate these rules. In addition, we present HREELS measurements of the phonon
dispersion for graphene on Ru(0001) and find excellent agreement with the theory. We also
illustrate the effect of different symmetry breaking scenarios in the selection rules and discuss
previous experiments in light of our results. 
\end{abstract}

\maketitle

%- The investigation of phonons in materials provides information on numerous
%physical properties, such as sound velocity, thermal expansion\cite{TVK81}, magnetic forces, heat
%capacity, and thermal conductivity. To measure the phonon spectrum of a material, many
%spectroscopic techniques can be used. 

\section{Introduction} The investigation of surface phonons is an invaluable tool to
study materials~\cite{KW91,BHR94}, as it provides a wealth of information on their
structural~\cite{PMR12}, electronic~\cite{PLM04}, magnetic~\cite{BHS01} or thermal properties
\cite{TVK81}, to name a few. Among many experimental probes, surface scattering experiments are
particularly well suited to measure phonon spectra. As in any scattering setup, however, the mapping
of the full
phonon dispersion with these methods is sometimes limited by selection rules~\cite{KW91,IM82}, which
preclude the observation of certain phonon branches. The understanding of these rules is therefore
crucial in the design and interpretation of these experiments. 

The origin of selection rules is the presence of symmetries that enforce conservation
laws. In surface scattering experiments, a selection rule applies when the scattering plane,
defined by the momenta of the incident and scattered particles, coincides with a mirror plane of the
surface. The selection rule states that phonons that are odd under this mirror reflection cannot be
observed~\cite{IM82,S85,BLI92,D90}, and can be easily understood as the conservation of parity under
reflections: since incoming and scattered wavefunctions of the probe have even parity, the
excitation of an odd parity phonon is forbidden, and the contribution to the cross section from this
process is zero. 

In experiments, the simplest geometry to measure phonon dispersions is planar scattering, with the
scattering plane perpendicular to the surface. Since one is usually interested in the
dispersion along high symmetry directions, the scattering plane is often a mirror plane and
selection rules apply. Knowledge of these rules can thus be of great help to interpret the data, for
example to assign symmetry labels to phonon branches or to detect broken symmetries. Moreover, this
understanding can be used to devise more complicated non-planar scattering geometries
~\cite{BLI92,YKW90,EJY90,GST96,D90} that are not affected by selection rules and allow one to
observe the odd modes~\footnote{A selection rule for a particular high-symmetry line can be avoided
altogether by choosing to measure $q$ in a replica of this line that does not map onto itself under
the corresponding reflection. This measurement usually requires two independent rotations of the
sample and is far less common.}.

To measure phonon dispersions, one of the most powerful experimental probes is high-resolution
electron energy loss spectroscopy (HREELS). Among other advantages, this technique offers excellent
energy and momentum resolution and allows one to map the full phonon spectrum. HREELS has been
applied
to many systems with great success, and is very useful in particular
to measure the spectrum of epitaxial monolayers grown on a substrate, where inelastic
neutron or X-ray scattering cannot be used. A well known example is the case of graphene
monolayers, where the effect of different substrates on the phonon spectrum has been widely studied
\cite{AHH92,SPR97,OAS88,A90,TAS95,FRS00,SFA99,ASO90,YTI05,ASI90,FSR99,HAH92,YTI04,AHO01}. 

\begin{figure*}[t]
\begin{center}
\includegraphics[width=18cm]{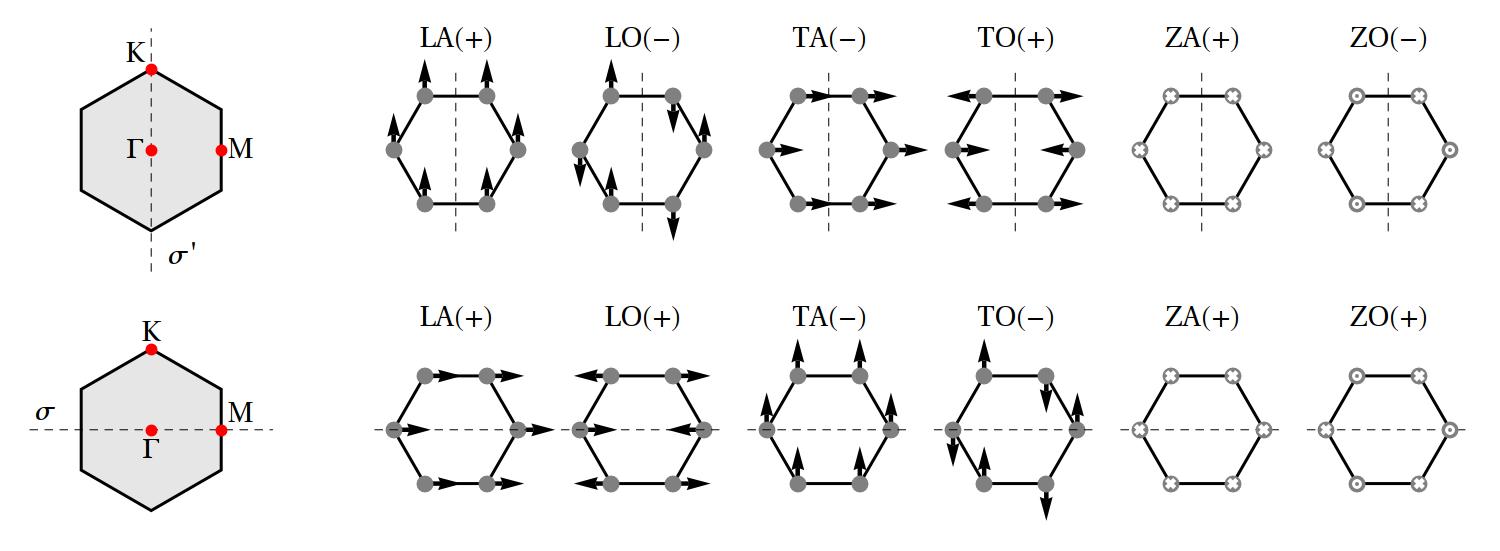}
\caption{ Honeycomb lattice phonon eigenvectors at the $\Gamma$ point, with their
polarization defined with respect to the $\Gamma K$ (top) and $\Gamma M$ (bottom) directions.  Out
of plane displacement is indicated by crosses (positive) and circles (negative). The mirror planes
$\sigma'$ (that leaves $\Gamma K$ invariant) and $\sigma$ (that leaves $\Gamma M$ invariant) are
represented as dashed lines. The parity of each phonon under the corresponding reflection is
indicated in parenthesis. Phonons with odd (-) parity are not observed in HREELS.
%a) The $\Gamma K$ direction in the Brillouin zone, which is invariant under the $\sigma'$
%mirror plane, denoted as a dashed line. b) Phonon eigenvectors at the $\Gamma$ point and their
%polarization with respect to the $\Gamma K$ direction.  Out of plane displacement is indicated by
%crosses (positive) and circles (negative). The parity eigenvalue with respect to $\sigma'$ is
%indicated in parenthesis. Phonons with odd (-) parity are not observed along this line. c-d) The
%same for the $\Gamma-M$ direction.
} \label{symm}
\end{center}
\end{figure*}

The case of HREELS studies of epitaxial graphene is of particular interest because, despite
the many experiments reported, their interpretation in terms of selection rules has often been
misleading. While most studies acknowledge the existence of a selection rule which forbids
the observation of the shear horizontal mode, SH, (or transverse acoustic, TA) along the $\Gamma M$
direction, other selection rules are sometimes misquoted and some have been completely missed. The
purpose of this work is to provide a detailed study of the selection rules for surfaces with
$C_{6v}$ symmetry, taking the case of graphene as an example. Our main result is the full set of
selection rules, summarized in Fig.~\ref{symm}: the modes TA and TO along the $\Gamma M$ direction
and the modes TA, ZO and LO along $\Gamma K$ are all odd and thus should not be observed. Our
results are worked out for HREELS for concreteness, but are equally applicable to any other planar
scattering experiment.

In the rest of this work, we first discuss how selection rules appear in the computation of
the HREELS scattering rate. We illustrate our results for this with a simplified phonon model, and
we compare them with our experimental HREELS data for graphene on Ru(0001). Finally, we will discuss
how symmetry breaking can render selection rules inactive, and interpret previous experiments in
light of our results. 

\section{Selection rules in the HREELS intensity} The origin of the selection rule explained in
the
introduction can be seen more explicitly by considering the computation of the HREELS
cross section~\cite{IM82,LTM80} due to phonon excitations. The relevant kinematic regime for this
process is known as impact scattering, where high-energy electrons interact with the short-range
part of the atomic potential. The incoming electron with energy $E_I$ and momentum ${\bf k}_I$ is
scattered off a surface and is recovered with energy $E_S$ and momentum ${\bf k}_S$. The excitation
of a phonon of frequency $\omega$ and momentum $\bf q$ is detected in the loss spectrum as a
resonance peak at $E_S = E_I \pm \omega$ and ${\bf k}_S = {\bf k}_I \pm \bf q$.

Because of the geometry of this problem it will be convenient to separate vectors into in-plane and
out of plane components, ${\bf q} = (\vec q_{\parallel}, q_z)$, reserving the arrow notation
$\vec q$ for two-dimensional vectors in the plane. For a given phonon of momentum $\vec
q_{\parallel}$ in the
Brillouin Zone (BZ) and eigenvector ${\bf u}^\alpha (q_\parallel)$, where $\alpha$ labels the
different atoms in the unit cell, the scattering amplitude in the impact scattering regime is
proportional to the matrix element~\cite{LTM80}
\begin{equation}\label{matel}
M =  \sum_{\alpha} e^{i \vec
q_{\parallel} \cdot \vec x_{\alpha}} {\bf q} \cdot {\bf u}^{\alpha} = \sum_{\alpha} e^{i \vec
q_{\parallel} \cdot \vec x_{\alpha}} \left[\vec q_{\parallel} \cdot \vec
u_{\parallel}^{\alpha} + q_z u_{z}^{\alpha}\right] ,
\end{equation}
where $\vec x_\alpha$ is the position of atom $\alpha$ in the unit cell. The scattering rate is
proportional to $|M|^2$. This matrix element accounts for the dominant changes in intensity as the
BZ is sampled for a fixed $E_I$. The full theory for the scattering cross section can be found in
Refs. \onlinecite{IM82,LTM80}, but knowledge of Eq.~\ref{matel} will suffice for our purposes. 

The selection rules can now be explained in terms of Eq.~\ref{matel}. When the scattering plane is
aligned with a mirror plane of the surface, ${\bf q}$ is invariant under the mirror plane. The
phonon eigenvectors at the corresponding $q_{\parallel}$ can thus be chosen with well defined parity
under this symmetry. The selection rule occurs because the matrix element must be invariant under
the symmetry, and since ${\bf q}$ is invariant, when ${\bf u}^\alpha (q_\parallel)$ is odd we must
have $M=-M$, which implies $M=0$. 

\section{Application to the honeycomb lattice} The honeycomb lattice has symmetry group $C_{6v}$,
which has two types of mirror planes, $\sigma$ and $\sigma'$ represented in Fig.~\ref{symm}. In the
BZ, the plane $\sigma'$ is aligned with the $\Gamma K$ direction, while the plane $\sigma$ is
aligned with $\Gamma M$. The selection rules are simple to state. For any surface with $C_{6v}$
symmetry and in a planar scattering experiment, if $q_{\parallel}$ lies in $\Gamma K$ odd phonons
under $\sigma'$ are not observed, while if $q_{\parallel}$ lies in $\Gamma M$ odd phonons under
$\sigma$ are not observed. The parity for any branch can be determined from its parity at $\Gamma$
in the absence of crossings. 

In the honeycomb lattice there are six phonon branches, four in-plane and two out-of-plane. At the
$\Gamma$ point, the in-plane acoustic (A) branches are degenerate and transform as $E_1$, while
the in-plane optical (O) branches transform as $E_2$. For these modes one can define transverse (T)
and longitudinal (L) polarizations along either $\Gamma K$ or $\Gamma M$, which determines
their transformation under reflections. The out-of-plane acoustical (ZA) and optical (ZO) transform
as $A_1$ and $B_2$ respectively. The parities under both reflections for all six branches are
illustrated in Fig.~\ref{symm} and are summarized as follows: the odd modes are the TA and TO in 
$\Gamma M$, and the the TA, LO and ZO in $\Gamma K$. The selection rule then states that none of
these modes should be observed in an HREELS experiment. 

\begin{figure}[t]
\begin{center}
\includegraphics[width=8.7cm]{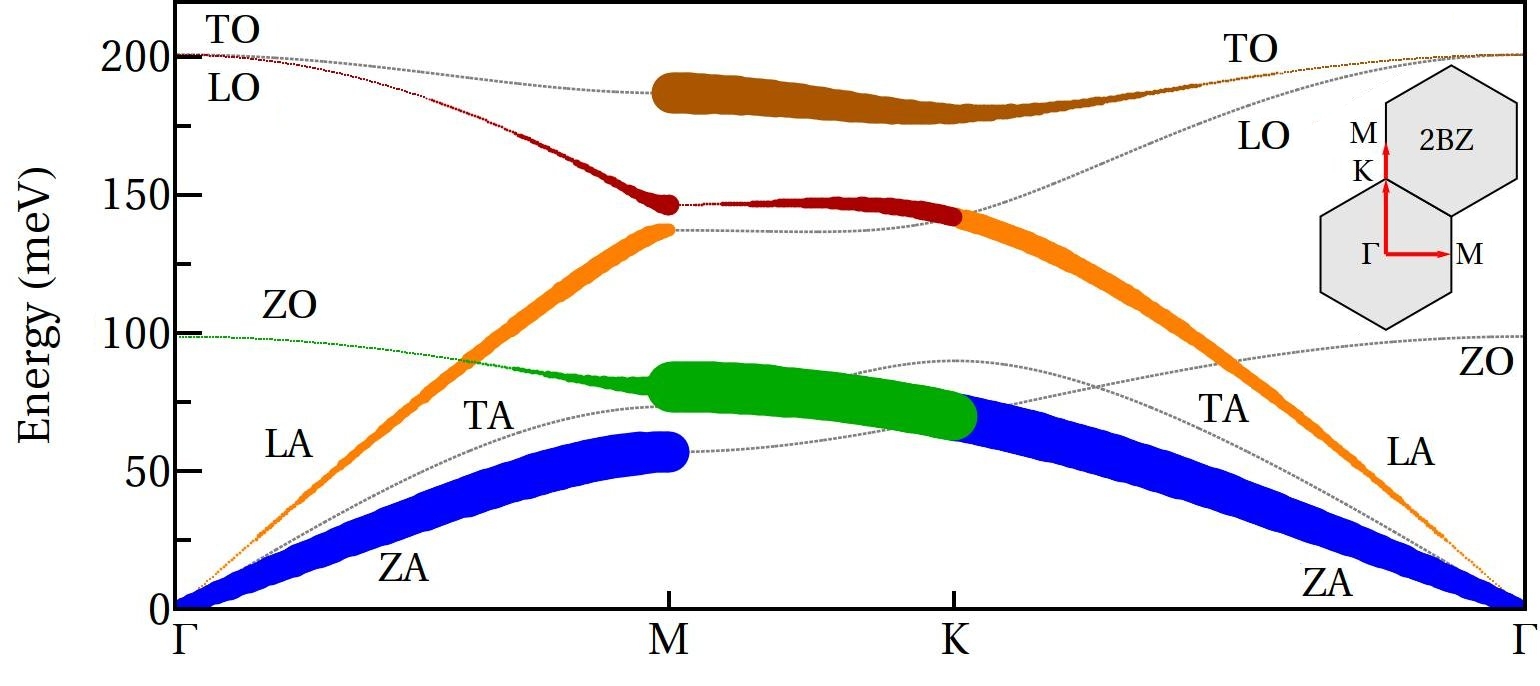}
\includegraphics[width=8.6cm]{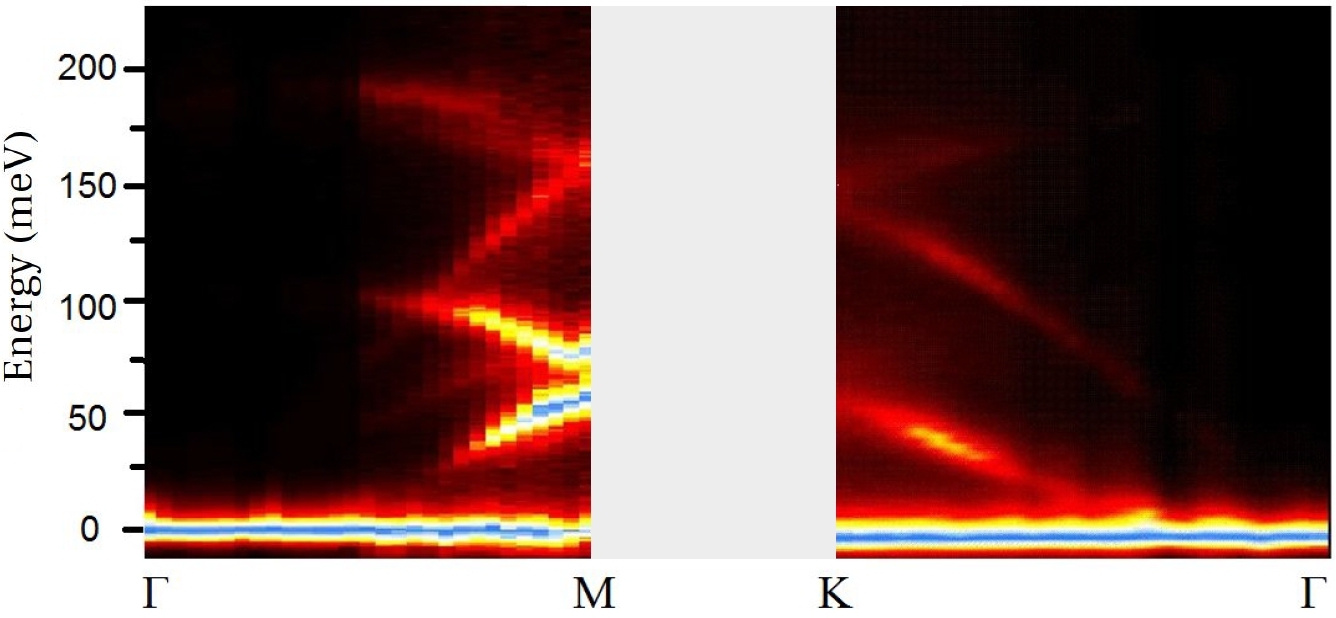}
\caption{a) Phonon spectrum for the model in Eqs.~\ref{inplane} and \ref{outplane} along the high
symmetry lines shown in the inset. The area of each dot is proportional to $|M|^2$.
Branches with zero intensity within numerical precision are shown as gray dotted lines. Note the
selection rules: TA and TO are absent in $\Gamma M$ and TA, ZO, LO are absent in $\Gamma K$. b)
Intensity map of the HREELS signal in planar scattering along the same symmetry lines. The agreement
for the selection rules of different branches is manifest.  Note that the uppermost observed branch
at $\Gamma K$ is the TO mode, which almost overlaps with the LO branch due to the softening induced
by a Kohn anomaly which is not captured by our simplified phonon model.}
\label{paths}
\end{center}
\end{figure}

This statement can be checked explicitly with a computation of the HREELS matrix element,
Eq.~\ref{matel}. Since our aim is to illustrate selection rules, we will use a minimal force
constant model to
describe the phonons, following Ref. \onlinecite{A90}. In this model the in-plane and out-of-plane
modes are decoupled and may be treated separately. For the in-plane modes the energy functional
includes nearest neighbor bond-stretching $\alpha_1$ and bond bending $\gamma_1$ terms. Denoting
the two sites in the unit cell $\alpha=A,B$, the energy is
\begin{align}
E &= \frac{\alpha_1}{2a^2}\sum_{\vec x, n} \left[\vec \delta_n \cdot (\vec u_{\vec x}^A -\vec
u_{\vec x+\vec \delta_n}^B)\right]^2 + \label{inplane}\\ & \frac{\gamma_1}{2a^2} \sum_{\vec x, n}
\left[ (\vec
u_{\vec x+\vec \delta_n}^A-\vec u_{\vec x}^B)\times \vec \delta_n -(\vec
u_{\vec x+\vec \delta_{n+1}}^A-\vec u_{\vec x}^B)\times \vec \delta_{n+1}\right]^2, \nonumber
\end{align}
where $\vec x = n\vec d_1 +m \vec d_2$ runs through all unit cells
and $\vec \delta_n$ is the nearest neighbor vector with $n=1,2,3$ and $a=|\vec \delta_n|$. The
out-of-plane modes are
modeled with an out-of-plane bond bending term $\gamma_2$
\begin{equation}
E = \frac{\gamma_2}{2} \sum_{\vec x, n} \left[u^A_{z,\vec x +\vec \delta_n}-u^B_{z,\vec
x}\right]^2.
\label{outplane}
\end{equation}
The phonon energies and eigenvectors are obtained from the equations of motion derived from
these energies, $m \omega^2 {\bf u}^{\alpha} = \partial E / \partial {\bf u}^{\alpha}$, with $m$ the
mass of a carbon atom, and the EELS matrix elements are then computed according to Eq.~\ref{matel}.
Note that for out-of-plane phonons of a given $q_{\parallel}$ the value of $q_z$ has to
be determined from kinematics from
\begin{equation}
q_z = 2 m_e E_{I} \left[1-\cos(\theta_I-\theta_S)-\frac{q_{\parallel}^2}{2mE_I}\right]^{1/2}
\end{equation}
where $E_I\sim E_S$ because $\omega_{ph}/E_S < 0.01$ for all phonon energies. The results of this
computation are shown in Fig.~\ref{paths} for the choice of parameters $\alpha_1=30$ $\rm eV
/\AA^{2}$, $\gamma_1=2.7$ $\rm eV /\AA^{2}$ and $\gamma_2=4.5$ $\rm eV /\AA^{2}$. The incident
energy is $E_I = 26$ eV. The different branches are plotted with point size
proportional to $|M|^2$ (with absolute scale chosen arbitrarily) and branches with zero intensity
are denoted with gray dotted lines. All selection rules are seen to apply as predicted.

\section{Comparison with experiment} To test our predictions for the selection rules, we have
measured high resolution EELS spectra of graphene
on Ru(0001)\cite{MGW07,MWB08,VCB08,PZS09,BBG10,MBS10,MWB10,PBM11,GDW11,AGJ12,PCF13}. This system is
an ideal playground to test symmetries of phonon modes since single domain samples extending up to
a millimeter square can be obtained, as revealed by STM studies \cite{PZS09}. This is in contrast
with other graphene/metal interfaces for which many domains with different azimuthal rotations are
found \cite{LNT09,CCC13}. The experimental phonon dispersion obtained is reported in
Fig.~\ref{paths}b. Data was taken for both the $\Gamma K$ and $\Gamma M$ directions in scattering
conditions enhancing the cross-section for phonon modes in graphitic systems, that is, primary
electron beam energy $E_I =$20 eV and grazing incidence. Other primary energies provide the same
phonon dispersion with reduced intensity. Notice that for $E_I =$20 eV, geometrical constraints of
the HREELS analyzer do not allow one to reach values of $q_{\parallel}$ sufficient to span the $M K$
line\footnote{The phonon dispersion along $M K$ was recorded with $E_I =$ 32 eV. However,
the cross-section of phonon modes in this scattering conditions became so weak that a
comparison with data acquired by using $E_I=$ 20 eV would not be meaningful.}. For the $\Gamma M$
and $\Gamma K$ lines measured in this work, the absence of the branches affected by selection rules
is manifest, and it is matched by the predictions: TA and TO are absent in $\Gamma M$ and TA,
ZO, LO are absent in $\Gamma K$. The general intensity trend is also correct, showing in particular
that there is higher intensity for the out of plane branches because $|q_z|>|q_{\parallel}|$. 

\begin{figure}[t]
\begin{center}
\includegraphics[width=8.5cm]{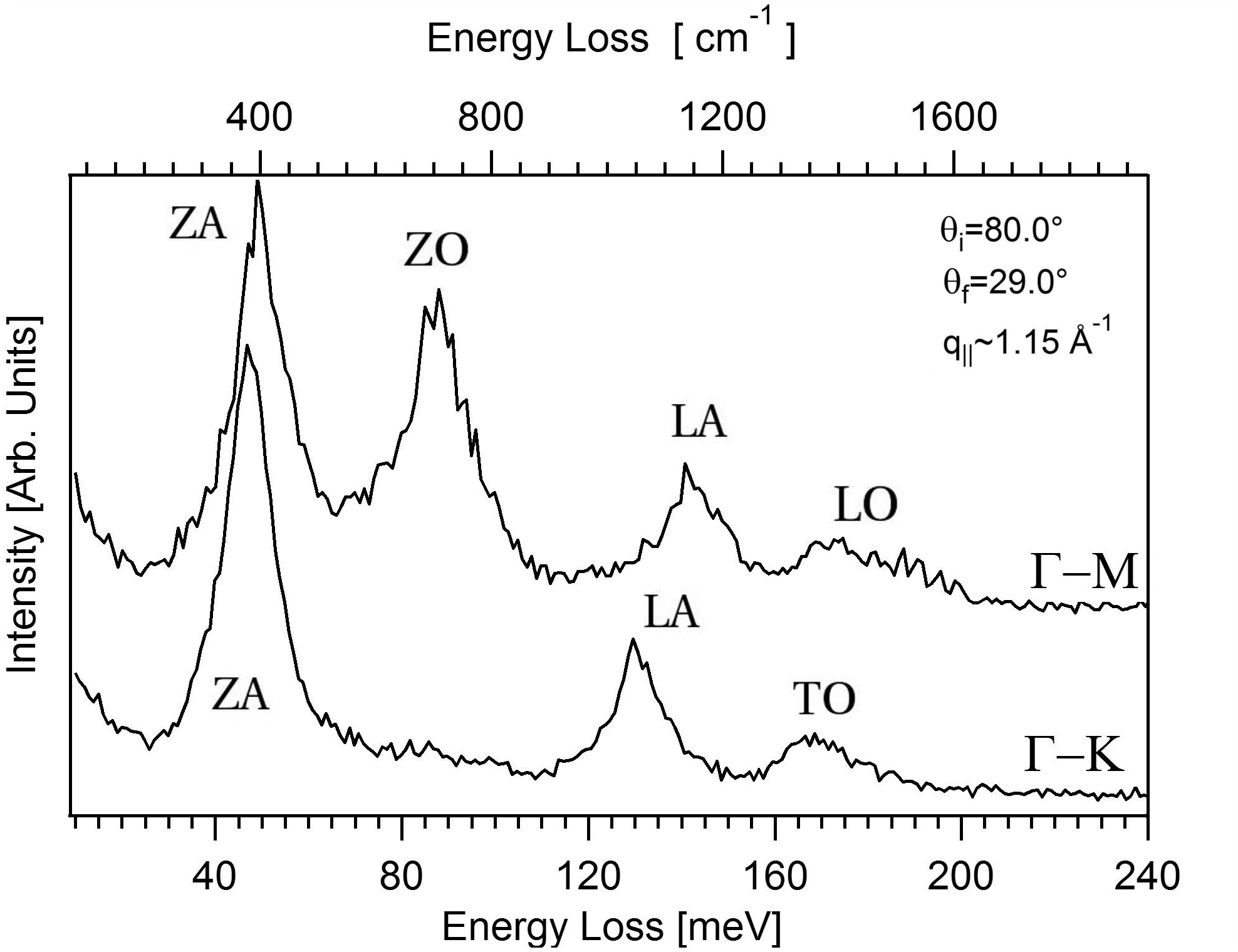}
\caption{HREEL spectra for momentum transfer $q_{\parallel} = 1.15 \rm \AA^{-1}$ in the
$\Gamma M$ and $\Gamma K$ directions. Note that the ZO
peak is missing in $\Gamma K$ due to the selection rule, and that the highest energy peak
corresponds to different phonons in the two different directions.} \label{ZOloss}
\end{center}
\end{figure}

The selection rule for the ZO in the $\Gamma K$ direction has not been noticed so far, but here
its effects are demonstrated clearly, as the ZO branch completely vanishes in this direction. This
effect is illustrated more explicitly in Fig.~\ref{ZOloss} where the EELS spectra for the $\Gamma K$
and $\Gamma M$ directions and $q_{\parallel} = 1.15 \rm \AA^{-1}$ are shown as a function of energy.
The absence of the ZO peak in $\Gamma K$ is clearly seen. Fig.~\ref{ZOloss} also emphasizes that
in $\Gamma K$, the highest frequency mode that is observed should be labeled TO, contrary to what
happens in $\Gamma M$, where the highest mode observed is LO. 

The phonon dispersion itself also matches reasonably well the experiment, and deviations only occur
for the branches affected by Kohn anomalies \cite{PLM04}, a well known limitation of short-ranged
models\cite{F08,VCC09} that do not account for the coupling to the $\pi$ electrons\cite{WR04,LAW08}.
The modes strongly affected by Kohn anomalies are the TO mode at $K$, which is shifted down almost
to the LO/LA crossing, and the LO at $\Gamma$, which should disperse upward faster than the TO, an
effect known as overbending. These two effects are missed by our model but can be reproduced with
ab-initio calculations including electron-phonon interactions \cite{LAW08}.

%\begin{figure}[h]
%\begin{center}
%\includegraphics[width=6.5cm]{final2.jpg}
%\includegraphics[width=6.5cm]{final2.jpg}
%\includegraphics[width=6.5cm]{final2.jpg}
%\caption{Breaking of selection rules.}
%\label{breaking}
%\end{center}
%\end{figure}

\section{Breaking selection rules} When an experiment reports finite but small intensity coming
from forbidden branches, it is important to understand the origin of this effect, as this may
provide information on the symmetries of the surface.  A first reason why forbidden branches can be
observable is simply the finite angle resolution of the detector, which will collect electrons in a
momentum range that necessarily goes out of the high symmetry line, introducing some contribution to
the intensity. For HREELS experiments this contribution is however negligible. Misalignment of the
high symmetry directions will also render selection rules inactive \cite{TAS95}. A second mechanism
is disorder, which
breaks translational symmetry. If momentum is not conserved, electrons collected at a particular
angle may have scattered phonons with a distribution of momenta for which the selection rule does
not apply. This may also happen for samples with good crystalline order but with domains of random
orientation. Thus when selection rules are violated this is commonly interpreted as a signal of
the disorder in the sample. 

\begin{figure*}[t]
\begin{center}
\includegraphics[width=8.7cm]{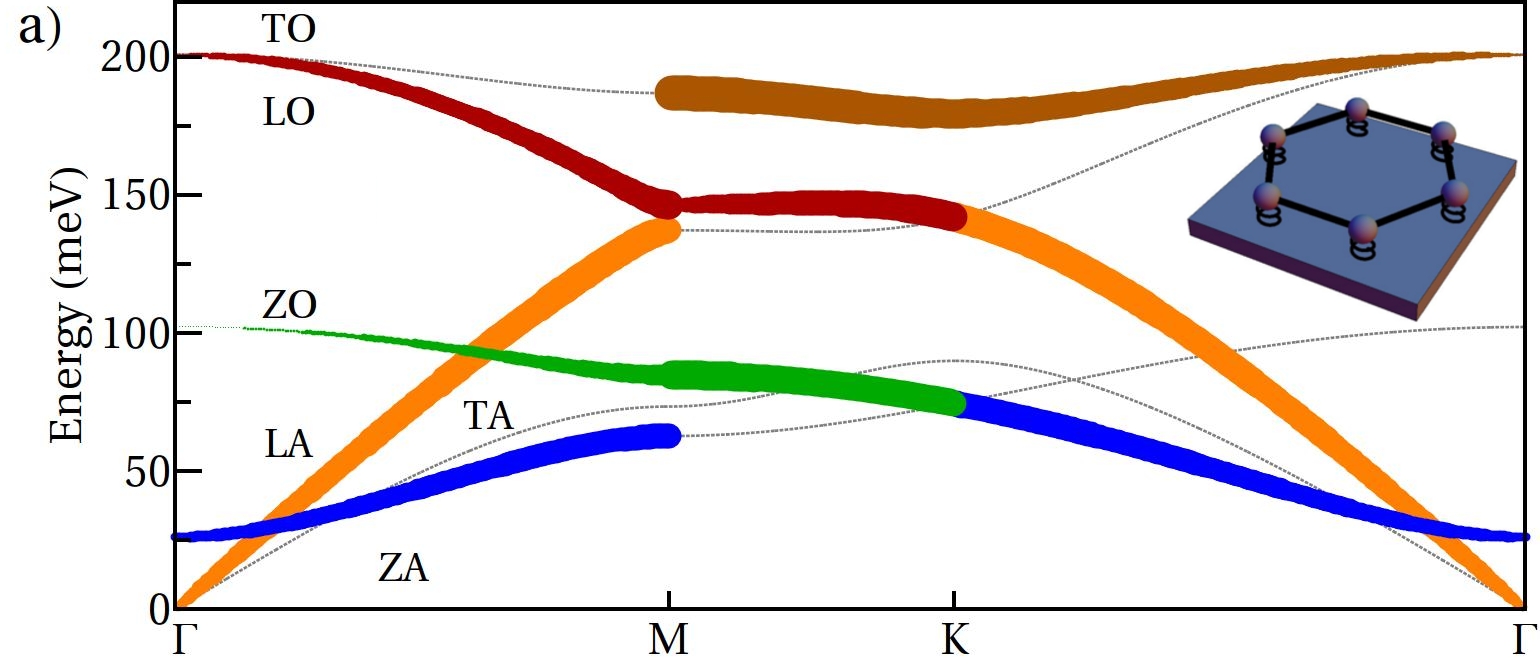}
\includegraphics[width=8.7cm]{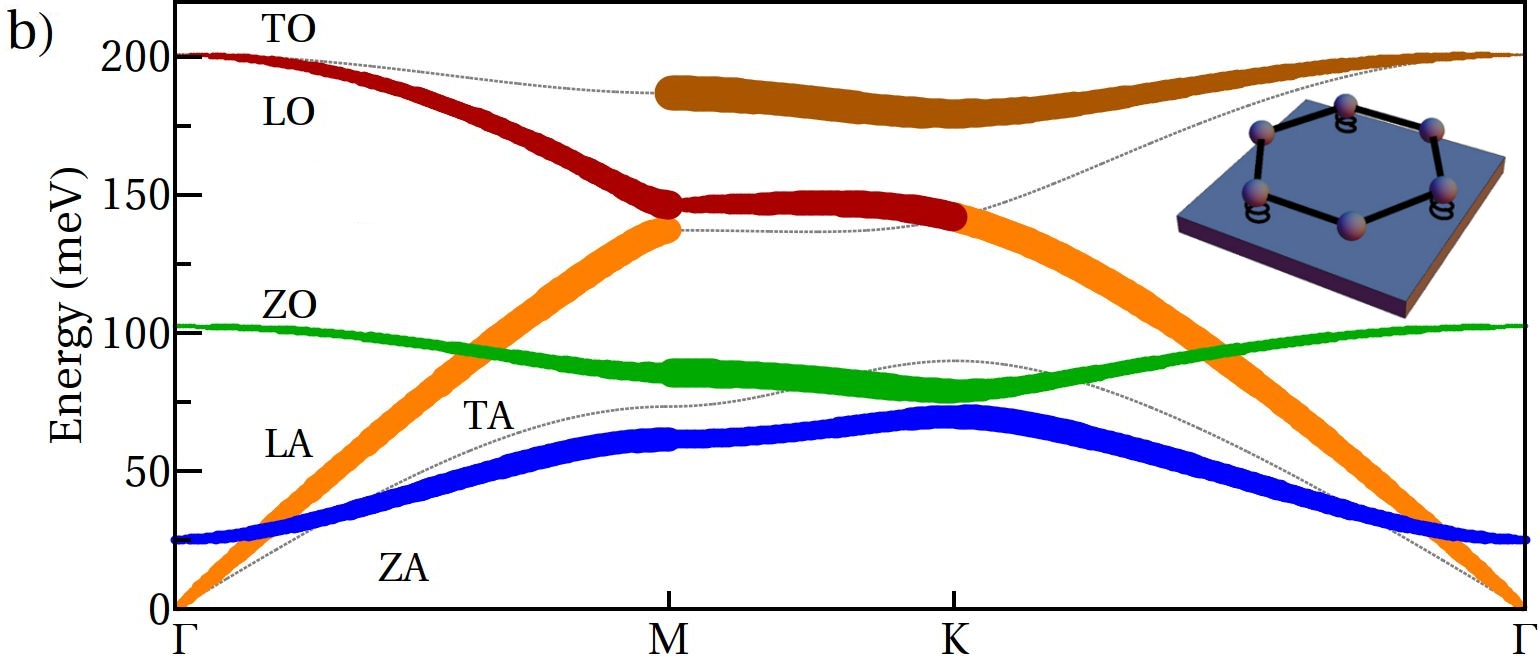}
\includegraphics[width=8.7cm]{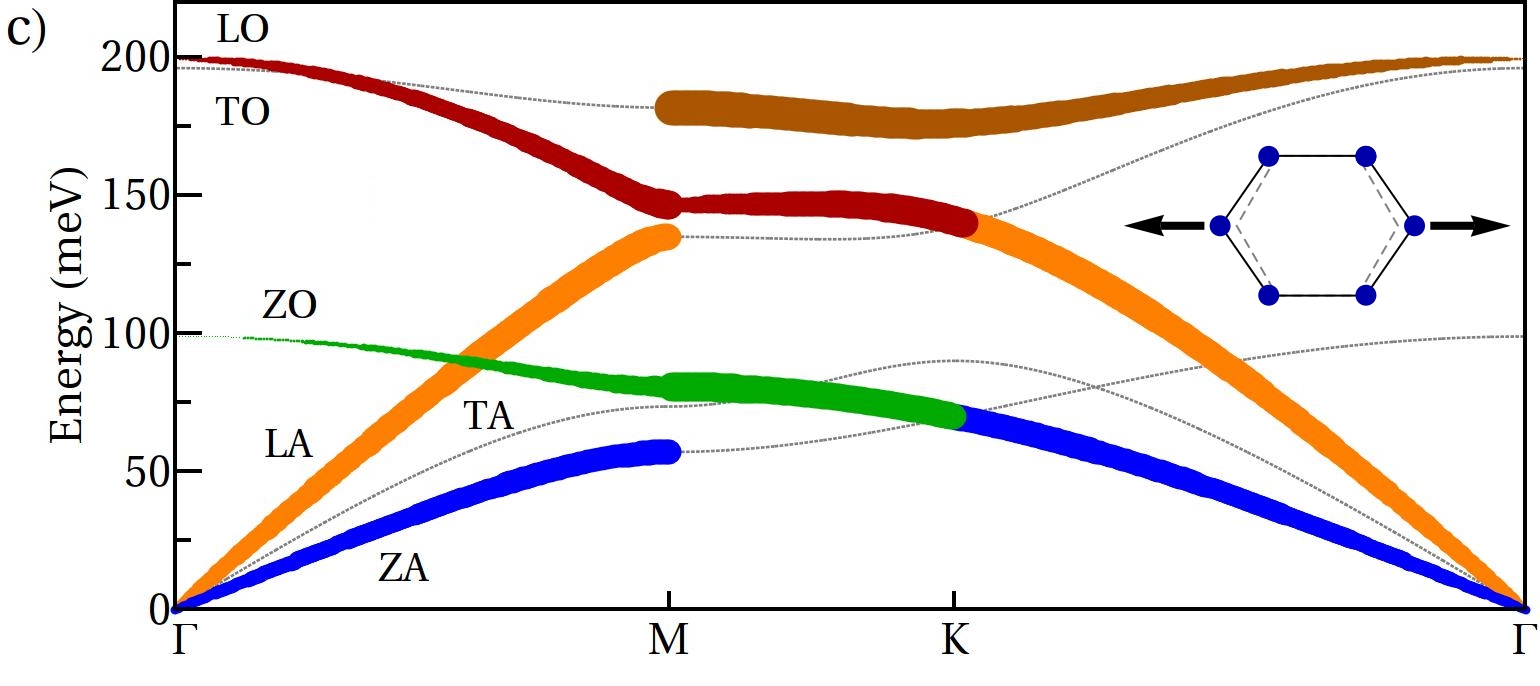}
\includegraphics[width=8.7cm]{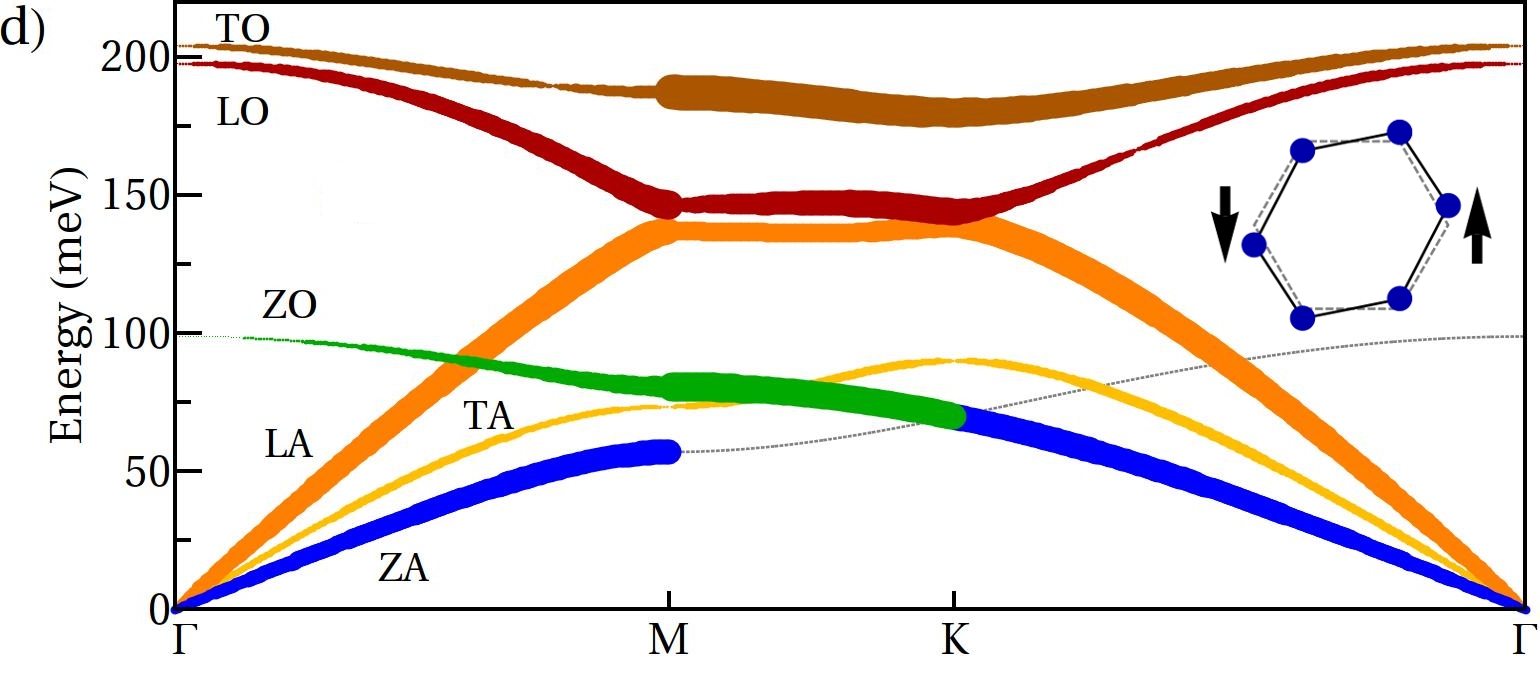}
\caption{EELS matrix element for different perturbations. Area of the dots is proportional to
$|M|^{1/2}$ for in-plane modes and to $0.2 |M|^{1/2}$ for out-of-plane modes so that most and less
intense modes (ZA and TA) can be seen in the same plot. a) A homogeneous coupling to the substrate
$\alpha_A = \alpha_B = 2 {\rm eV /\AA^{2}}$  lifts the ZA mode but breaks no symmetry, so
selection rules are preserved. b) A coupling to only one sublattice $\alpha_A = 4 {\rm eV /\AA^{2}}$
breaks the $\sigma'$ reflection and renders $\Gamma K$ selection rules inactive. c) Uniaxial strain
$u_{xx}=0.01$ splits the LO/TO crossing at $\Gamma$, but again breaks no reflections. d) Shear
strain $u_{xy} =0.01$ breaks both reflections and renders all selection rules for in-plane modes
inactive.} \label{breaking2}
\end{center}
\end{figure*}

When the previous mechanisms have been excluded, the violation of a selection rule is likely
the result of symmetry breaking. For example, the surface under study may be reconstructed with
reduced symmetry. Or in our case, the substrate where the graphene monolayer is grown may introduce
strain or additional spring constants which again break the symmetry $C_{6v}$ to a lower subgroup. 
We now explore this last case in more detail, considering
two natural examples. First, consider a honeycomb lattice that lies on top of a substrate. This can
be modeled by spring constants $\alpha_A,\alpha_B$ connecting each sublattice to the substrate
\begin{equation}
E_{\rm substrate} = \frac{1}{2}\sum_{\vec x} \alpha_A (u^A_z)^2 + \alpha_B (u^B_z)^2.
\end{equation}
These couplings have typical values of 1-5 $\rm eV/\AA^{2}$ for different
metals\cite{A90,AW10}. For a homogeneous substrate we have $\alpha_A=\alpha_B$, while for a
perfectly commensurate triangular lattice substrate, for example for graphene on Ni(111), we have
$\alpha_A\neq0$, $\alpha_B=0$.  In the latter case, the point is group is reduced to $C_{3v}$ and
the $\sigma'$ reflection is broken, so that selection rules in $\Gamma K$ become inactive for
out-of-plane modes. The computed EELS intensity for these two types of substrates is shown in
Figs.~\ref{breaking2}a-b. While both perturbations affect the spectrum, we see that only the second
one changes the selection rules, making the $ZO$ visible in $\Gamma K$. Since the substrate coupling
only affects $u_z$, selection rules for in-plane modes remain valid. 

In the second case, we consider a sample under constant strain $u_{ij}$, which could be either
substrate induced or externally applied. The uniaxial components of strain $u_{xx}$,$u_{yy}$ break
the
point group symmetry to $C_{2v}$ (which still has a $\sigma$ and $\sigma'$ reflection) while the
shear component $u_{xy}$ breaks it to $C_2$ which has no reflections. Therefore, shear strain
will remove selection rules completely. Strain can be introduced in our model by modifying the
bond stretching couplings. If the relative change in nearest neighbor distance due to strain is
 $\Delta u_n = (|\vec \delta'_n| - |\vec \delta_n|)/a \approx \delta_n^i \delta_n^j u_{ij}/a^2$, the
change in the bond stretching constant can be parametrized as $\alpha_{1,n} = \alpha_1(1-\beta
\Delta u_n)$, with $\beta = |\partial \log \alpha_1/\partial \log a |$. In this model $\beta$ is
related to the Gruneisen parameter as $\beta = 4 \gamma_{E_{2g}}$, with $\gamma_{E_{2g}}\approx2$
as estimated from the Raman splitting of the G mode under strain \cite{MLN09}. The bond
stretching energy in the presence of strain is modified to
\begin{align}
E_{\rm strain} &= \frac{1}{2a^2}\sum_{\vec x, n} \alpha_{1,n}\left[\vec \delta_n \cdot (\vec u_x^A
-\vec u_{x+\delta_n}^B)\right]^2.  \label{strain}
\end{align}
The EELS matrix element for these two types of strains is shown in Figs.~\ref{breaking2}c-d. Again,
both perturbations change the spectrum, but only the second one changes the intensity pattern: since
shear strain breaks both reflections, all in-plane modes become visible.

\section{Discussion} One of the conclusions of this work is that the selection rules usually
quoted
in the literature have been often incomplete or misunderstood. The confusion may originate from the
fact that in the EELS literature, surface phonons are classified into two groups: sagittal plane
(SP) phonons have polarization parallel to the sagittal (or scattering) plane, while shear
horizontal
(SH) phonons have polarization perpendicular to this plane~\cite{KW91}. This classification has
sometimes lead to a formulation of selection rules that states that SH modes are not observed in
planar scattering. This formulation can be misleading because
it implicitly assumes that the polarization behaves as a vector under reflections. When this is the
case, SH modes are indeed odd under reflection, while SP modes are even. For example, this happens
for the in-plane acoustic modes which transform as an $E_1$ representation. One may thus identify
the LA as SP and the TA as SH, which is not observed. However, this formulation is incorrect for an
arbitrary representation, for example for the optical in-plane phonons (LO,TO), which transform as
$E_2$. In the $\Gamma K$ direction, the TO has polarization perpendicular to the scattering plane
and is thus labeled as SHO~\cite{AHH92,OAS88}, SH*~\cite{SPR97,SFR98,FSR99,FRS00} or simply
SH~\cite{A90,ASO90}. Nevertheless it is even under reflection and has no selection rule, contrary
to common belief \cite{SPR97}. In the same way, the LO is SP (parallel to the plane) but is odd
under reflection and should be absent. When only one of the LO/TO phonons is observed, the label
that should be assigned to it is thus TO (or SH).

A similar situation occurs for the ZO (a $B_2$ representation) in the $\Gamma K$ direction, which is
polarized in the $z$ direction (also SP) but is odd under reflection. This selection rule has
been missed until now because this mode is not a shear mode, but as Fig.~\ref{ZOloss} clearly shows,
this mode is not observed on graphene on Ru(0001) in $\Gamma K$, but is visibly observed in
$\Gamma M$. This is also consistent with intensity data available in the literature. The ZO in
$\Gamma K$ is almost invisible in graphene on Ni(111) when intercalated with Ag\cite{FSR99}, but is
clearly visible along $\Gamma M$ on graphene on BC$_3$ \cite{YTI05}.

Our analysis of selection rules also sheds light on previous EELS experiments performed on graphite.
The surface phonons of graphite in the (001) direction are very similar to those of graphene because
of the weak interlayer coupling. However, the planar symmetry group is actually $C_{3v}$ because
the graphene layers are stacked in an alternating AB sequence: only one sublattice has carbon atoms
below. As a result, the $\sigma'$ reflection is broken and there are no
selection rules in the $\Gamma K$ direction. The only remaining selection rules apply to the TA and
TO phonons in the $\Gamma M$ direction. The same analysis would apply to any related material
with $C_{3v}$ symmetry such as BN\cite{RHS97} or silicene\cite{VDQ12,FFR12}.

The EELS spectra of graphite in Ref.~\onlinecite{WPW87} were recorded in a sample containing
domains with random azimuthal direction. For this reason, the TA mode was clearly observed.
However, in Ref.~\onlinecite{OAS88} both TO and TA were observed in the $\Gamma M$ direction, which
is inconsistent with the selection rule. As noted in Ref.~\onlinecite{WR04}, in this experiment TA
and ZO
cross before reaching the $M$ point, which is also
inconsistent with both theory calculations and with more recent X-ray data\cite{MMD07} where there
is a clear TA/ZO crossing at $M$. Since the crossing in Ref.~\onlinecite{OAS88} occurs approximately
at the
same $q$ as the crossing in the $\Gamma K$ and the selection rule is being violated, it appears
possible that the measurement may have contained several orientational domains as well. It would be
interesting to see new experimental data to shed light on this issue.

In summary, in this work we have provided the selection rules for the measurement of the phonon
spectrum of the honeycomb lattice in planar scattering, showing that some selection rules have been
overlooked. These are however manifest in our HREELS spectra in graphene on Ru(0001). We hope that
our work will serve as a guideline for further experiments measuring phonon dispersions in graphene
or other materials. 

\begin{acknowledgments}
The authors would like to thank Davide Campi for helpful discussions. F. de J. acknowledges support 
from the ``Programa Nacional de Movilidad de Recursos Humanos" (Spanish MECD). This work was
supported in part by the NSF through Grant No. DMR-1005035 and by the US-Israel Binational Science
Foundation.
\end{acknowledgments}

\section{Appendix}
\subsection{Experimental methods}

Experiments were carried out in an ultra-high vacuum (UHV) chamber
operating at a base pressure of 5$\cdot$10$^{-11}$ mbar. The
sample was a single crystal of Ru(0001) which was cleaned
by repeated cycles of ion sputtering and annealing at 1300 K.
Surface cleanliness and order were checked using Auger electron
spectroscopy (AES) and low-energy electron diffraction (LEED)
measurements, respectively.

\begin{figure}[h]
\begin{center}
\includegraphics[width=6cm]{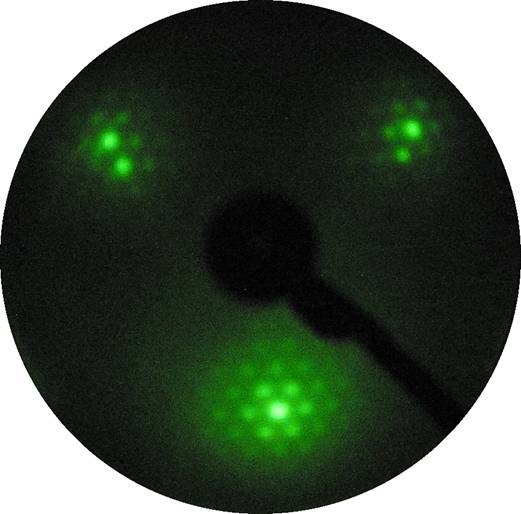}
\caption{LEED pattern of graphene on Ru(0001), recorded at $E_p$ = 74 eV and room temperature.}
\label{LEED}
\end{center}
\end{figure}

Graphene was obtained by dosing ethylene onto the clean Ru(0001) substrate held at 1150 K. The MLG
was reached upon an exposure of 3$\cdot$10$^{-8}$ mbar for ten minutes (24 L. 1
L=1.33$\cdot$10$^{-6}$ mbar·s). After removing the C$_2$H$_4$ gas from the chamber the temperature
was held at 1150 K for further 60 seconds.
The attained LEED pattern (shown in Fig.~\ref{LEED}) is essentially
similar to those previously reported \cite{MGW07,PZS09}. Around each spot of the (1x1), additional
spots due to the (12x12) reconstruction of the overlayer were revealed. Only MLG has been
observed in the whole sample, as in STM and He atom scattering experiments by Politano et al.
reported elsewhere \cite{BBG10,PBM11}.

HREELS experiments were performed by using an
electron energy loss spectrometer (Delta 0.5, SPECS). The energy
resolution of the spectrometer was degraded to 4 meV so as to
increase the signal-to-noise ratio of loss peaks. Dispersion of
the loss peaks, i.e., E$_{loss}$(q$_{||}$), was measured by moving
the analyzer while keeping the sample and the monochromator in a
fixed position. To measure the dispersion relation, values for the
parameters E$_{I}$, impinging energy and $\theta_{I}$, the
incident angle, were chosen so as to obtain the highest
signal-to-noise ratio. The primary beam energy used for the
dispersion, E$_{I}$=20 eV, provided, in fact, the best compromise
among surface sensitivity, the highest cross-section for mode
excitation and q$_{||}$ resolution. As 
\begin{displaymath}
 {q_{\vert \vert}}
 =
 \hbar\left(k_{I} \sin\theta _{I} - k_{S} \sin\theta _{S}
 \right),
\end{displaymath}
the parallel momentum transfer $q_{\vert\vert }$ depends on
E$_{I}$, E$_{loss}$, $\theta _{I}$ and $\theta _{S}$ according to
\begin{displaymath}
 {q}_{\vert \vert}
 =
 \frac{\sqrt{2mE_{I}}}{\hbar}
 \left(
 \sin\theta _{\textrm{I}}-\sqrt{1-\frac{{E}_{loss}}{E_{I}}} \sin\theta _{S}
 \right)
\end{displaymath}
where E$_{loss}$ is the energy loss and $\theta _{S}$ is the
electron scattering angle \cite{R95}. Accordingly, the integration window in reciprocal space is
\cite{PFC09}
\begin{displaymath}
 {\Delta q}_{\vert \vert}
 \approx
 \frac{\sqrt{2mE_{I}}}{\hbar}
 \left(
 \cos\theta _{I}+\sqrt{1-\frac{{E}_{loss}}{E_{I}}} \cos\theta _{S}
 \right)\cdot\alpha
\end{displaymath}
where $\alpha$ is the angular acceptance of the apparatus \cite{PFC08}
($\pm$0.5\r{} in our case). For the investigated range of
q$_{||}$, the indeterminacy has been found to range from 0.005
(near $\Gamma$) to 0.022 $\AA^{-1}$ (for higher momenta). To
obtain the intensities of phonon modes, a polynomial background
was subtracted from each spectrum. All
measurements were made at room temperature.

\subsection{Appendix B: Strain and Gruneisen parameter}

The effect of strain was described in the main text as a change in the bond stretching constant,
which becomes neighbor dependent $\alpha_{1,n} = \alpha_1(1-\beta \delta_n^i \delta_n^j
u_{ij}/a^2)$. The parameter $\beta$ is related to the Gruneisen parameter for
the optical phonon at $\Gamma$, defined as\cite{MHC09,MLN09}
\begin{equation}
\gamma_{E_{2g}} = \frac{1}{\omega_{E_{2g}}} \frac{\partial \omega_{E_{2g}}}{\partial u_h},
\end{equation}
where $u_h = u_{xx} + u_{yy}$. To show this relation, consider the total energy in the presence of
strain
\begin{align}
E &= \frac{1}{2a^2}\sum_{\vec x, n} \alpha_1(1-\beta \delta_n^i \delta_n^j
u_{ij}/a^2) \left[\vec \delta_n \cdot (\vec u_x^A -\vec
u_{x+\delta_n}^B)\right]^2 + \nonumber \\ & \frac{\gamma_1}{2a^2} \sum_{\vec x, n} \left[ (\vec
u_{x+\delta_n}^A-\vec u_x^B)\times \vec \delta_n -(\vec
u_{x+\delta_{n+1}}^A-\vec u_x^B)\times \vec \delta_{n+1}\right]^2,
\end{align}
where we assume that $\gamma_1$ does not change. The equation of motion is
\begin{align}
M (\omega^2 - \omega_{E_{2g}}^2) u_i = - 3\alpha_1 \frac{\beta}{4}( u_h \; \mathcal{I}_{ij}+
2u_{ij}) u_j,
\end{align}
with $\omega_{E_{2g}}^2 = (3\alpha_1 + 9 \gamma_1)/M$ the frequency of the optical mode, and we
have set $\vec u^A=-\vec u^B=\vec u$ for the optical mode. Neglecting $\gamma_1$ compared with
$\alpha_1$, specifying to uniaxial strain $u_{xy}=0$ and taking the square root for small strain we
obtain
\begin{equation}
\omega_{\pm} \approx \omega_{E_{2g}} -  \omega_{E_{2g}} \frac{
\beta}{8}(2 u_h \pm (u_{xx}-u_{yy})),
\end{equation}
which gives $\gamma_{E_{2g}} = \beta/4$.

%As a consequence of the existence of ripples in monolayer graphene on
%Ru(0001)\cite{VCB08,PZS09,BBG10,MBS10,MWB10,PBM11,AGJ12,PCF13}, the low-energy electron diffraction
%(LEED) pattern shows satellite spots caused by the moiré structure (Fig. S1). For a microscopic
%characterization by scanning tunnelling microscopy (STM), see Refs \cite{MGW07,MWB08,GDW11} .

%\subsection{Equations of motion}
%The equation of motion is
%\begin{align}
%M\omega^2 u_x^i =& \left(\frac{3\alpha_1}{2}+\frac{9\gamma_1}{2}\right) u_x^i \nonumber \\-& \sum_n
%\left(\alpha_1\delta_n^i\delta_n^j + 3\gamma_1
%\epsilon^{ik}\epsilon^{jl}\delta_n^k\delta_n^l\right)u^i_{x+\delta_n}
%\end{align}

%and the equation of motion is
%\begin{equation}
%M \omega^2 u^z_x= 3 \gamma_2 ( 3 u^z_x -\sum_{\vec x, n}u^z_{x +\delta_n} )
%\end{equation}

\bibliography{rules}

\begin{thebibliography}{59}%
\makeatletter
\providecommand \@ifxundefined [1]{%
 \@ifx{#1\undefined}
}%
\providecommand \@ifnum [1]{%
 \ifnum #1\expandafter \@firstoftwo
 \else \expandafter \@secondoftwo
 \fi
}%
\providecommand \@ifx [1]{%
 \ifx #1\expandafter \@firstoftwo
 \else \expandafter \@secondoftwo
 \fi
}%
\providecommand \natexlab [1]{#1}%
\providecommand \enquote  [1]{``#1''}%
\providecommand \bibnamefont  [1]{#1}%
\providecommand \bibfnamefont [1]{#1}%
\providecommand \citenamefont [1]{#1}%
\providecommand \href@noop [0]{\@secondoftwo}%
\providecommand \href [0]{\begingroup \@sanitize@url \@href}%
\providecommand \@href[1]{\@@startlink{#1}\@@href}%
\providecommand \@@href[1]{\endgroup#1\@@endlink}%
\providecommand \@sanitize@url [0]{\catcode `\\12\catcode `\$12\catcode
  `\&12\catcode `\#12\catcode `\^12\catcode `\_12\catcode `\%12\relax}%
\providecommand \@@startlink[1]{}%
\providecommand \@@endlink[0]{}%
\providecommand \url  [0]{\begingroup\@sanitize@url \@url }%
\providecommand \@url [1]{\endgroup\@href {#1}{\urlprefix }}%
\providecommand \urlprefix  [0]{URL }%
\providecommand \Eprint [0]{\href }%
\providecommand \doibase [0]{http://dx.doi.org/}%
\providecommand \selectlanguage [0]{\@gobble}%
\providecommand \bibinfo  [0]{\@secondoftwo}%
\providecommand \bibfield  [0]{\@secondoftwo}%
\providecommand \translation [1]{[#1]}%
\providecommand \BibitemOpen [0]{}%
\providecommand \bibitemStop [0]{}%
\providecommand \bibitemNoStop [0]{.\EOS\space}%
\providecommand \EOS [0]{\spacefactor3000\relax}%
\providecommand \BibitemShut  [1]{\csname bibitem#1\endcsname}%
\let\auto@bib@innerbib\@empty
%</preamble>
\bibitem [{\citenamefont {Kress}\ and\ \citenamefont {de~Wette}()}]{KW91}%
  \BibitemOpen
  \bibfield  {author} {\bibinfo {author} {\bibfnamefont {W.}~\bibnamefont
  {Kress}}\ and\ \bibinfo {author} {\bibfnamefont {F.~W.}\ \bibnamefont
  {de~Wette}},\ }\href@noop {} {\emph {\bibinfo {title} {Surface Phonons}}}\
  (\bibinfo  {publisher} {Springer})\BibitemShut {NoStop}%
\bibitem [{\citenamefont {Benedek}\ \emph {et~al.}(1994)\citenamefont
  {Benedek}, \citenamefont {Hofmann}, \citenamefont {Ruggerone}, \citenamefont
  {Onida},\ and\ \citenamefont {Miglio}}]{BHR94}%
  \BibitemOpen
  \bibfield  {author} {\bibinfo {author} {\bibfnamefont {G.}~\bibnamefont
  {Benedek}}, \bibinfo {author} {\bibfnamefont {F.}~\bibnamefont {Hofmann}},
  \bibinfo {author} {\bibfnamefont {P.}~\bibnamefont {Ruggerone}}, \bibinfo
  {author} {\bibfnamefont {G.}~\bibnamefont {Onida}}, \ and\ \bibinfo {author}
  {\bibfnamefont {L.}~\bibnamefont {Miglio}},\ }\href@noop {} {\bibfield
  {journal} {\bibinfo  {journal} {Surf. Sci. Rep.}\ }\textbf {\bibinfo {volume}
  {20}},\ \bibinfo {pages} {1} (\bibinfo {year} {1994})}\BibitemShut {NoStop}%
\bibitem [{\citenamefont {Politano}\ \emph {et~al.}(2012)\citenamefont
  {Politano}, \citenamefont {Marino}, \citenamefont {Campi}, \citenamefont
  {Far{\'\i}as}, \citenamefont {Miranda},\ and\ \citenamefont
  {Chiarello}}]{PMR12}%
  \BibitemOpen
  \bibfield  {author} {\bibinfo {author} {\bibfnamefont {A.}~\bibnamefont
  {Politano}}, \bibinfo {author} {\bibfnamefont {A.~R.}\ \bibnamefont
  {Marino}}, \bibinfo {author} {\bibfnamefont {D.}~\bibnamefont {Campi}},
  \bibinfo {author} {\bibfnamefont {D.}~\bibnamefont {Far{\'\i}as}}, \bibinfo
  {author} {\bibfnamefont {R.}~\bibnamefont {Miranda}}, \ and\ \bibinfo
  {author} {\bibfnamefont {G.}~\bibnamefont {Chiarello}},\ }\href@noop {}
  {\bibfield  {journal} {\bibinfo  {journal} {Carbon}\ }\textbf {\bibinfo
  {volume} {50}},\ \bibinfo {pages} {4903} (\bibinfo {year}
  {2012})}\BibitemShut {NoStop}%
\bibitem [{\citenamefont {Piscanec}\ \emph {et~al.}(2004)\citenamefont
  {Piscanec}, \citenamefont {Lazzeri}, \citenamefont {Mauri}, \citenamefont
  {Ferrari},\ and\ \citenamefont {Robertson}}]{PLM04}%
  \BibitemOpen
  \bibfield  {author} {\bibinfo {author} {\bibfnamefont {S.}~\bibnamefont
  {Piscanec}}, \bibinfo {author} {\bibfnamefont {M.}~\bibnamefont {Lazzeri}},
  \bibinfo {author} {\bibfnamefont {F.}~\bibnamefont {Mauri}}, \bibinfo
  {author} {\bibfnamefont {A.~C.}\ \bibnamefont {Ferrari}}, \ and\ \bibinfo
  {author} {\bibfnamefont {J.}~\bibnamefont {Robertson}},\ }\href {\doibase
  10.1103/PhysRevLett.93.185503} {\bibfield  {journal} {\bibinfo  {journal}
  {Phys. Rev. Lett.}\ }\textbf {\bibinfo {volume} {93}},\ \bibinfo {pages}
  {185503} (\bibinfo {year} {2004})}\BibitemShut {NoStop}%
\bibitem [{\citenamefont {Benedek}\ \emph {et~al.}(2001)\citenamefont
  {Benedek}, \citenamefont {Hulpke},\ and\ \citenamefont
  {Steinh\"ogl}}]{BHS01}%
  \BibitemOpen
  \bibfield  {author} {\bibinfo {author} {\bibfnamefont {G.}~\bibnamefont
  {Benedek}}, \bibinfo {author} {\bibfnamefont {E.}~\bibnamefont {Hulpke}}, \
  and\ \bibinfo {author} {\bibfnamefont {W.}~\bibnamefont {Steinh\"ogl}},\
  }\href {\doibase 10.1103/PhysRevLett.87.027201} {\bibfield  {journal}
  {\bibinfo  {journal} {Phys. Rev. Lett.}\ }\textbf {\bibinfo {volume} {87}},\
  \bibinfo {pages} {027201} (\bibinfo {year} {2001})}\BibitemShut {NoStop}%
\bibitem [{\citenamefont {Talwar}\ \emph {et~al.}(1981)\citenamefont {Talwar},
  \citenamefont {Vandevyver}, \citenamefont {Kunc},\ and\ \citenamefont
  {Zigone}}]{TVK81}%
  \BibitemOpen
  \bibfield  {author} {\bibinfo {author} {\bibfnamefont {D.~N.}\ \bibnamefont
  {Talwar}}, \bibinfo {author} {\bibfnamefont {M.}~\bibnamefont {Vandevyver}},
  \bibinfo {author} {\bibfnamefont {K.}~\bibnamefont {Kunc}}, \ and\ \bibinfo
  {author} {\bibfnamefont {M.}~\bibnamefont {Zigone}},\ }\href {\doibase
  10.1103/PhysRevB.24.741} {\bibfield  {journal} {\bibinfo  {journal} {Phys.
  Rev. B}\ }\textbf {\bibinfo {volume} {24}},\ \bibinfo {pages} {741} (\bibinfo
  {year} {1981})}\BibitemShut {NoStop}%
\bibitem [{\citenamefont {Ibach}\ and\ \citenamefont {Mills}()}]{IM82}%
  \BibitemOpen
  \bibfield  {author} {\bibinfo {author} {\bibfnamefont {H.}~\bibnamefont
  {Ibach}}\ and\ \bibinfo {author} {\bibfnamefont {D.}~\bibnamefont {Mills}},\
  }\href@noop {} {\emph {\bibinfo {title} {Electron Energy Loss Spectroscopy
  and Surface Vibrations}}}\ (\bibinfo  {publisher} {Academic
  Press})\BibitemShut {NoStop}%
\bibitem [{\citenamefont {Szeftel}(1985)}]{S85}%
  \BibitemOpen
  \bibfield  {author} {\bibinfo {author} {\bibfnamefont {J.}~\bibnamefont
  {Szeftel}},\ }\href {\doibase http://dx.doi.org/10.1016/0039-6028(85)90490-X}
  {\bibfield  {journal} {\bibinfo  {journal} {Surf. Sci.}\ }\textbf {\bibinfo
  {volume} {152}},\ \bibinfo {pages} {797 } (\bibinfo {year}
  {1985})}\BibitemShut {NoStop}%
\bibitem [{\citenamefont {Balden}\ \emph {et~al.}(1992)\citenamefont {Balden},
  \citenamefont {Lehwald}, \citenamefont {Ibach}, \citenamefont {Ormeci},\ and\
  \citenamefont {Mills}}]{BLI92}%
  \BibitemOpen
  \bibfield  {author} {\bibinfo {author} {\bibfnamefont {M.}~\bibnamefont
  {Balden}}, \bibinfo {author} {\bibfnamefont {S.}~\bibnamefont {Lehwald}},
  \bibinfo {author} {\bibfnamefont {H.}~\bibnamefont {Ibach}}, \bibinfo
  {author} {\bibfnamefont {A.}~\bibnamefont {Ormeci}}, \ and\ \bibinfo {author}
  {\bibfnamefont {D.~L.}\ \bibnamefont {Mills}},\ }\href {\doibase
  10.1103/PhysRevB.46.4172} {\bibfield  {journal} {\bibinfo  {journal} {Phys.
  Rev. B}\ }\textbf {\bibinfo {volume} {46}},\ \bibinfo {pages} {4172}
  (\bibinfo {year} {1992})}\BibitemShut {NoStop}%
\bibitem [{\citenamefont {Doak}(1990)}]{D90}%
  \BibitemOpen
  \bibfield  {author} {\bibinfo {author} {\bibfnamefont {R.}~\bibnamefont
  {Doak}},\ }\href {\doibase 10.1016/0749-6036(90)90295-I} {\bibfield
  {journal} {\bibinfo  {journal} {Superlattices Microstruct.}\ }\textbf
  {\bibinfo {volume} {7}},\ \bibinfo {pages} {201 } (\bibinfo {year}
  {1990})}\BibitemShut {NoStop}%
\bibitem [{\citenamefont {Yater}\ \emph {et~al.}(1990)\citenamefont {Yater},
  \citenamefont {Kulkarni}, \citenamefont {de~Wette},\ and\ \citenamefont
  {Erskine}}]{YKW90}%
  \BibitemOpen
  \bibfield  {author} {\bibinfo {author} {\bibfnamefont {J.}~\bibnamefont
  {Yater}}, \bibinfo {author} {\bibfnamefont {A.}~\bibnamefont {Kulkarni}},
  \bibinfo {author} {\bibfnamefont {F.}~\bibnamefont {de~Wette}}, \ and\
  \bibinfo {author} {\bibfnamefont {J.}~\bibnamefont {Erskine}},\ }\href
  {\doibase 10.1016/0368-2048(90)80232-Y} {\bibfield  {journal} {\bibinfo
  {journal} {J. Electron Spectros. Rel. Phenom.}\ }\textbf {\bibinfo {volume}
  {54}},\ \bibinfo {pages} {395 } (\bibinfo {year} {1990})}\BibitemShut
  {NoStop}%
\bibitem [{\citenamefont {Erskine}\ \emph {et~al.}(1990)\citenamefont
  {Erskine}, \citenamefont {Jeong}, \citenamefont {Yater}, \citenamefont
  {Chen},\ and\ \citenamefont {Tong}}]{EJY90}%
  \BibitemOpen
  \bibfield  {author} {\bibinfo {author} {\bibfnamefont {J.~L.}\ \bibnamefont
  {Erskine}}, \bibinfo {author} {\bibfnamefont {E.}~\bibnamefont {Jeong}},
  \bibinfo {author} {\bibfnamefont {J.}~\bibnamefont {Yater}}, \bibinfo
  {author} {\bibfnamefont {Y.}~\bibnamefont {Chen}}, \ and\ \bibinfo {author}
  {\bibfnamefont {S.~Y.}\ \bibnamefont {Tong}},\ }\href {\doibase
  10.1116/1.576687} {\bibfield  {journal} {\bibinfo  {journal} {J. Vac. Sci.
  Technol. A}\ }\textbf {\bibinfo {volume} {8}},\ \bibinfo {pages} {2649}
  (\bibinfo {year} {1990})}\BibitemShut {NoStop}%
\bibitem [{\citenamefont {Glebov}\ \emph {et~al.}(1996)\citenamefont {Glebov},
  \citenamefont {Silvestri}, \citenamefont {Toennies}, \citenamefont
  {Benedek},\ and\ \citenamefont {Skofronick}}]{GST96}%
  \BibitemOpen
  \bibfield  {author} {\bibinfo {author} {\bibfnamefont {A.}~\bibnamefont
  {Glebov}}, \bibinfo {author} {\bibfnamefont {W.}~\bibnamefont {Silvestri}},
  \bibinfo {author} {\bibfnamefont {J.~P.}\ \bibnamefont {Toennies}}, \bibinfo
  {author} {\bibfnamefont {G.}~\bibnamefont {Benedek}}, \ and\ \bibinfo
  {author} {\bibfnamefont {J.~G.}\ \bibnamefont {Skofronick}},\ }\href
  {\doibase 10.1103/PhysRevB.54.17866} {\bibfield  {journal} {\bibinfo
  {journal} {Phys. Rev. B}\ }\textbf {\bibinfo {volume} {54}},\ \bibinfo
  {pages} {17866} (\bibinfo {year} {1996})}\BibitemShut {NoStop}%
\bibitem [{Note1()}]{Note1}%
  \BibitemOpen
  \bibinfo {note} {A selection rule for a particular high-symmetry line can be
  avoided altogether by choosing to measure $q$ in a replica of this line that
  does not map onto itself under the corresponding reflection. This measurement
  usually requires two independent rotations of the sample and is far less
  common.}\BibitemShut {Stop}%
\bibitem [{\citenamefont {Aizawa}\ \emph {et~al.}(1992)\citenamefont {Aizawa},
  \citenamefont {Hwang}, \citenamefont {Hayami}, \citenamefont {Souda},
  \citenamefont {Otani},\ and\ \citenamefont {Ishizawa}}]{AHH92}%
  \BibitemOpen
  \bibfield  {author} {\bibinfo {author} {\bibfnamefont {T.}~\bibnamefont
  {Aizawa}}, \bibinfo {author} {\bibfnamefont {Y.}~\bibnamefont {Hwang}},
  \bibinfo {author} {\bibfnamefont {W.}~\bibnamefont {Hayami}}, \bibinfo
  {author} {\bibfnamefont {R.}~\bibnamefont {Souda}}, \bibinfo {author}
  {\bibfnamefont {S.}~\bibnamefont {Otani}}, \ and\ \bibinfo {author}
  {\bibfnamefont {Y.}~\bibnamefont {Ishizawa}},\ }\href {\doibase
  http://dx.doi.org/10.1016/0039-6028(92)90046-9} {\bibfield  {journal}
  {\bibinfo  {journal} {Surf. Sci.}\ }\textbf {\bibinfo {volume} {260}},\
  \bibinfo {pages} {311 } (\bibinfo {year} {1992})}\BibitemShut {NoStop}%
\bibitem [{\citenamefont {Siebentritt}\ \emph {et~al.}(1997)\citenamefont
  {Siebentritt}, \citenamefont {Pues}, \citenamefont {Rieder},\ and\
  \citenamefont {Shikin}}]{SPR97}%
  \BibitemOpen
  \bibfield  {author} {\bibinfo {author} {\bibfnamefont {S.}~\bibnamefont
  {Siebentritt}}, \bibinfo {author} {\bibfnamefont {R.}~\bibnamefont {Pues}},
  \bibinfo {author} {\bibfnamefont {K.-H.}\ \bibnamefont {Rieder}}, \ and\
  \bibinfo {author} {\bibfnamefont {A.~M.}\ \bibnamefont {Shikin}},\ }\href
  {\doibase 10.1103/PhysRevB.55.7927} {\bibfield  {journal} {\bibinfo
  {journal} {Phys. Rev. B}\ }\textbf {\bibinfo {volume} {55}},\ \bibinfo
  {pages} {7927} (\bibinfo {year} {1997})}\BibitemShut {NoStop}%
\bibitem [{\citenamefont {Oshima}\ \emph {et~al.}(1988)\citenamefont {Oshima},
  \citenamefont {Aizawa}, \citenamefont {Souda}, \citenamefont {Ishizawa},\
  and\ \citenamefont {Sumiyoshi}}]{OAS88}%
  \BibitemOpen
  \bibfield  {author} {\bibinfo {author} {\bibfnamefont {C.}~\bibnamefont
  {Oshima}}, \bibinfo {author} {\bibfnamefont {T.}~\bibnamefont {Aizawa}},
  \bibinfo {author} {\bibfnamefont {R.}~\bibnamefont {Souda}}, \bibinfo
  {author} {\bibfnamefont {Y.}~\bibnamefont {Ishizawa}}, \ and\ \bibinfo
  {author} {\bibfnamefont {Y.}~\bibnamefont {Sumiyoshi}},\ }\href@noop {}
  {\bibfield  {journal} {\bibinfo  {journal} {Solid State Commun.}\ }\textbf
  {\bibinfo {volume} {65}},\ \bibinfo {pages} {1601} (\bibinfo {year}
  {1988})}\BibitemShut {NoStop}%
\bibitem [{\citenamefont {Aizawa}\ \emph
  {et~al.}(1990{\natexlab{a}})\citenamefont {Aizawa}, \citenamefont {Souda},
  \citenamefont {Otani}, \citenamefont {Ishizawa},\ and\ \citenamefont
  {Oshima}}]{A90}%
  \BibitemOpen
  \bibfield  {author} {\bibinfo {author} {\bibfnamefont {T.}~\bibnamefont
  {Aizawa}}, \bibinfo {author} {\bibfnamefont {R.}~\bibnamefont {Souda}},
  \bibinfo {author} {\bibfnamefont {S.}~\bibnamefont {Otani}}, \bibinfo
  {author} {\bibfnamefont {Y.}~\bibnamefont {Ishizawa}}, \ and\ \bibinfo
  {author} {\bibfnamefont {C.}~\bibnamefont {Oshima}},\ }\href {\doibase
  10.1103/PhysRevB.42.11469} {\bibfield  {journal} {\bibinfo  {journal} {Phys.
  Rev. B}\ }\textbf {\bibinfo {volume} {42}},\ \bibinfo {pages} {11469}
  (\bibinfo {year} {1990}{\natexlab{a}})}\BibitemShut {NoStop}%
\bibitem [{\citenamefont {Tilley}\ \emph {et~al.}(1995)\citenamefont {Tilley},
  \citenamefont {Aizawa}, \citenamefont {Souda}, \citenamefont {Hayami},
  \citenamefont {Otani},\ and\ \citenamefont {Ishizawa}}]{TAS95}%
  \BibitemOpen
  \bibfield  {author} {\bibinfo {author} {\bibfnamefont {B.}~\bibnamefont
  {Tilley}}, \bibinfo {author} {\bibfnamefont {T.}~\bibnamefont {Aizawa}},
  \bibinfo {author} {\bibfnamefont {R.}~\bibnamefont {Souda}}, \bibinfo
  {author} {\bibfnamefont {W.}~\bibnamefont {Hayami}}, \bibinfo {author}
  {\bibfnamefont {S.}~\bibnamefont {Otani}}, \ and\ \bibinfo {author}
  {\bibfnamefont {Y.}~\bibnamefont {Ishizawa}},\ }\href@noop {} {\bibfield
  {journal} {\bibinfo  {journal} {Solid State Commun.}\ }\textbf {\bibinfo
  {volume} {94}},\ \bibinfo {pages} {685} (\bibinfo {year} {1995})}\BibitemShut
  {NoStop}%
\bibitem [{\citenamefont {Far\'ias}\ \emph {et~al.}(2000)\citenamefont
  {Far\'ias}, \citenamefont {Rieder}, \citenamefont {Shikin}, \citenamefont
  {Adamchuk}, \citenamefont {Tanaka},\ and\ \citenamefont {Oshima}}]{FRS00}%
  \BibitemOpen
  \bibfield  {author} {\bibinfo {author} {\bibfnamefont {D.}~\bibnamefont
  {Far\'ias}}, \bibinfo {author} {\bibfnamefont {K.}~\bibnamefont {Rieder}},
  \bibinfo {author} {\bibfnamefont {A.}~\bibnamefont {Shikin}}, \bibinfo
  {author} {\bibfnamefont {V.}~\bibnamefont {Adamchuk}}, \bibinfo {author}
  {\bibfnamefont {T.}~\bibnamefont {Tanaka}}, \ and\ \bibinfo {author}
  {\bibfnamefont {C.}~\bibnamefont {Oshima}},\ }\href {\doibase
  10.1016/S0039-6028(00)00253-3} {\bibfield  {journal} {\bibinfo  {journal}
  {Surf. Sci.}\ }\textbf {\bibinfo {volume} {454}},\ \bibinfo {pages} {437 }
  (\bibinfo {year} {2000})}\BibitemShut {NoStop}%
\bibitem [{\citenamefont {Shikin}\ \emph {et~al.}(1999)\citenamefont {Shikin},
  \citenamefont {Far\'ias}, \citenamefont {Adamchuk},\ and\ \citenamefont
  {Rieder}}]{SFA99}%
  \BibitemOpen
  \bibfield  {author} {\bibinfo {author} {\bibfnamefont {A.}~\bibnamefont
  {Shikin}}, \bibinfo {author} {\bibfnamefont {D.}~\bibnamefont {Far\'ias}},
  \bibinfo {author} {\bibfnamefont {V.}~\bibnamefont {Adamchuk}}, \ and\
  \bibinfo {author} {\bibfnamefont {K.-H.}\ \bibnamefont {Rieder}},\ }\href
  {\doibase 10.1016/S0039-6028(99)00099-0} {\bibfield  {journal} {\bibinfo
  {journal} {Surf. Sci.}\ }\textbf {\bibinfo {volume} {424}},\ \bibinfo {pages}
  {155 } (\bibinfo {year} {1999})}\BibitemShut {NoStop}%
\bibitem [{\citenamefont {Aizawa}\ \emph
  {et~al.}(1990{\natexlab{b}})\citenamefont {Aizawa}, \citenamefont {Souda},
  \citenamefont {Otani}, \citenamefont {Ishizawa},\ and\ \citenamefont
  {Oshima}}]{ASO90}%
  \BibitemOpen
  \bibfield  {author} {\bibinfo {author} {\bibfnamefont {T.}~\bibnamefont
  {Aizawa}}, \bibinfo {author} {\bibfnamefont {R.}~\bibnamefont {Souda}},
  \bibinfo {author} {\bibfnamefont {S.}~\bibnamefont {Otani}}, \bibinfo
  {author} {\bibfnamefont {Y.}~\bibnamefont {Ishizawa}}, \ and\ \bibinfo
  {author} {\bibfnamefont {C.}~\bibnamefont {Oshima}},\ }\href {\doibase
  10.1103/PhysRevLett.64.768} {\bibfield  {journal} {\bibinfo  {journal} {Phys.
  Rev. Lett.}\ }\textbf {\bibinfo {volume} {64}},\ \bibinfo {pages} {768}
  (\bibinfo {year} {1990}{\natexlab{b}})}\BibitemShut {NoStop}%
\bibitem [{\citenamefont {Yanagisawa}\ \emph {et~al.}(2005)\citenamefont
  {Yanagisawa}, \citenamefont {Tanaka}, \citenamefont {Ishida}, \citenamefont
  {Matsue}, \citenamefont {Rokuta}, \citenamefont {Otani},\ and\ \citenamefont
  {Oshima}}]{YTI05}%
  \BibitemOpen
  \bibfield  {author} {\bibinfo {author} {\bibfnamefont {H.}~\bibnamefont
  {Yanagisawa}}, \bibinfo {author} {\bibfnamefont {T.}~\bibnamefont {Tanaka}},
  \bibinfo {author} {\bibfnamefont {Y.}~\bibnamefont {Ishida}}, \bibinfo
  {author} {\bibfnamefont {M.}~\bibnamefont {Matsue}}, \bibinfo {author}
  {\bibfnamefont {E.}~\bibnamefont {Rokuta}}, \bibinfo {author} {\bibfnamefont
  {S.}~\bibnamefont {Otani}}, \ and\ \bibinfo {author} {\bibfnamefont
  {C.}~\bibnamefont {Oshima}},\ }\href {\doibase 10.1002/sia.1948} {\bibfield
  {journal} {\bibinfo  {journal} {Surf. Interface Anal.}\ }\textbf {\bibinfo
  {volume} {37}},\ \bibinfo {pages} {133} (\bibinfo {year} {2005})}\BibitemShut
  {NoStop}%
\bibitem [{\citenamefont {Aizawa}\ \emph
  {et~al.}(1990{\natexlab{c}})\citenamefont {Aizawa}, \citenamefont {Souda},
  \citenamefont {Ishizawa}, \citenamefont {Hirano}, \citenamefont {Yamada},
  \citenamefont {ichi Tanaka},\ and\ \citenamefont {Oshima}}]{ASI90}%
  \BibitemOpen
  \bibfield  {author} {\bibinfo {author} {\bibfnamefont {T.}~\bibnamefont
  {Aizawa}}, \bibinfo {author} {\bibfnamefont {R.}~\bibnamefont {Souda}},
  \bibinfo {author} {\bibfnamefont {Y.}~\bibnamefont {Ishizawa}}, \bibinfo
  {author} {\bibfnamefont {H.}~\bibnamefont {Hirano}}, \bibinfo {author}
  {\bibfnamefont {T.}~\bibnamefont {Yamada}}, \bibinfo {author} {\bibfnamefont
  {K.}~\bibnamefont {ichi Tanaka}}, \ and\ \bibinfo {author} {\bibfnamefont
  {C.}~\bibnamefont {Oshima}},\ }\href {\doibase 10.1016/0039-6028(90)90531-C}
  {\bibfield  {journal} {\bibinfo  {journal} {Surf. Sci.}\ }\textbf {\bibinfo
  {volume} {237}},\ \bibinfo {pages} {194 } (\bibinfo {year}
  {1990}{\natexlab{c}})}\BibitemShut {NoStop}%
\bibitem [{\citenamefont {Far\'ias}\ \emph {et~al.}(1999)\citenamefont
  {Far\'ias}, \citenamefont {Shikin}, \citenamefont {Rieder},\ and\
  \citenamefont {Dedkov}}]{FSR99}%
  \BibitemOpen
  \bibfield  {author} {\bibinfo {author} {\bibfnamefont {D.}~\bibnamefont
  {Far\'ias}}, \bibinfo {author} {\bibfnamefont {A.~M.}\ \bibnamefont
  {Shikin}}, \bibinfo {author} {\bibfnamefont {K.-H.}\ \bibnamefont {Rieder}},
  \ and\ \bibinfo {author} {\bibfnamefont {Y.~S.}\ \bibnamefont {Dedkov}},\
  }\href@noop {} {\bibfield  {journal} {\bibinfo  {journal} {J. Phys.: Condens.
  Matter}\ }\textbf {\bibinfo {volume} {11}},\ \bibinfo {pages} {8453}
  (\bibinfo {year} {1999})}\BibitemShut {NoStop}%
\bibitem [{\citenamefont {Hwang}\ \emph {et~al.}(1992)\citenamefont {Hwang},
  \citenamefont {Aizawa}, \citenamefont {Hayami}, \citenamefont {Otani},
  \citenamefont {Ishizawa},\ and\ \citenamefont {Park}}]{HAH92}%
  \BibitemOpen
  \bibfield  {author} {\bibinfo {author} {\bibfnamefont {Y.}~\bibnamefont
  {Hwang}}, \bibinfo {author} {\bibfnamefont {T.}~\bibnamefont {Aizawa}},
  \bibinfo {author} {\bibfnamefont {W.}~\bibnamefont {Hayami}}, \bibinfo
  {author} {\bibfnamefont {S.}~\bibnamefont {Otani}}, \bibinfo {author}
  {\bibfnamefont {Y.}~\bibnamefont {Ishizawa}}, \ and\ \bibinfo {author}
  {\bibfnamefont {S.-J.}\ \bibnamefont {Park}},\ }\href {\doibase
  10.1016/0039-6028(92)90886-B} {\bibfield  {journal} {\bibinfo  {journal}
  {Surf. Sci.}\ }\textbf {\bibinfo {volume} {271}},\ \bibinfo {pages} {299 }
  (\bibinfo {year} {1992})}\BibitemShut {NoStop}%
\bibitem [{\citenamefont {Yanagisawa}\ \emph {et~al.}(2004)\citenamefont
  {Yanagisawa}, \citenamefont {Tanaka}, \citenamefont {Ishida}, \citenamefont
  {Matsue}, \citenamefont {Rokuta}, \citenamefont {Otani},\ and\ \citenamefont
  {Oshima}}]{YTI04}%
  \BibitemOpen
  \bibfield  {author} {\bibinfo {author} {\bibfnamefont {H.}~\bibnamefont
  {Yanagisawa}}, \bibinfo {author} {\bibfnamefont {T.}~\bibnamefont {Tanaka}},
  \bibinfo {author} {\bibfnamefont {Y.}~\bibnamefont {Ishida}}, \bibinfo
  {author} {\bibfnamefont {M.}~\bibnamefont {Matsue}}, \bibinfo {author}
  {\bibfnamefont {E.}~\bibnamefont {Rokuta}}, \bibinfo {author} {\bibfnamefont
  {S.}~\bibnamefont {Otani}}, \ and\ \bibinfo {author} {\bibfnamefont
  {C.}~\bibnamefont {Oshima}},\ }\href {\doibase 10.1103/PhysRevLett.93.177003}
  {\bibfield  {journal} {\bibinfo  {journal} {Phys. Rev. Lett.}\ }\textbf
  {\bibinfo {volume} {93}},\ \bibinfo {pages} {177003} (\bibinfo {year}
  {2004})}\BibitemShut {NoStop}%
\bibitem [{\citenamefont {Aizawa}\ \emph {et~al.}(2001)\citenamefont {Aizawa},
  \citenamefont {Hayami},\ and\ \citenamefont {Otani}}]{AHO01}%
  \BibitemOpen
  \bibfield  {author} {\bibinfo {author} {\bibfnamefont {T.}~\bibnamefont
  {Aizawa}}, \bibinfo {author} {\bibfnamefont {W.}~\bibnamefont {Hayami}}, \
  and\ \bibinfo {author} {\bibfnamefont {S.}~\bibnamefont {Otani}},\ }\href
  {\doibase 10.1103/PhysRevB.65.024303} {\bibfield  {journal} {\bibinfo
  {journal} {Phys. Rev. B}\ }\textbf {\bibinfo {volume} {65}},\ \bibinfo
  {pages} {024303} (\bibinfo {year} {2001})}\BibitemShut {NoStop}%
\bibitem [{\citenamefont {Li}\ \emph {et~al.}(1980)\citenamefont {Li},
  \citenamefont {Tong},\ and\ \citenamefont {Mills}}]{LTM80}%
  \BibitemOpen
  \bibfield  {author} {\bibinfo {author} {\bibfnamefont {C.~H.}\ \bibnamefont
  {Li}}, \bibinfo {author} {\bibfnamefont {S.~Y.}\ \bibnamefont {Tong}}, \ and\
  \bibinfo {author} {\bibfnamefont {D.~L.}\ \bibnamefont {Mills}},\ }\href
  {\doibase 10.1103/PhysRevB.21.3057} {\bibfield  {journal} {\bibinfo
  {journal} {Phys. Rev. B}\ }\textbf {\bibinfo {volume} {21}},\ \bibinfo
  {pages} {3057} (\bibinfo {year} {1980})}\BibitemShut {NoStop}%
\bibitem [{\citenamefont {Marchini}\ \emph {et~al.}(2007)\citenamefont
  {Marchini}, \citenamefont {G\"unther},\ and\ \citenamefont
  {Wintterlin}}]{MGW07}%
  \BibitemOpen
  \bibfield  {author} {\bibinfo {author} {\bibfnamefont {S.}~\bibnamefont
  {Marchini}}, \bibinfo {author} {\bibfnamefont {S.}~\bibnamefont {G\"unther}},
  \ and\ \bibinfo {author} {\bibfnamefont {J.}~\bibnamefont {Wintterlin}},\
  }\href {\doibase 10.1103/PhysRevB.76.075429} {\bibfield  {journal} {\bibinfo
  {journal} {Phys. Rev. B}\ }\textbf {\bibinfo {volume} {76}},\ \bibinfo
  {pages} {075429} (\bibinfo {year} {2007})}\BibitemShut {NoStop}%
\bibitem [{\citenamefont {Martoccia}\ \emph {et~al.}(2008)\citenamefont
  {Martoccia}, \citenamefont {Willmott}, \citenamefont {Brugger}, \citenamefont
  {Bj\"orck}, \citenamefont {G\"unther}, \citenamefont {Schlep\"utz},
  \citenamefont {Cervellino}, \citenamefont {Pauli}, \citenamefont {Patterson},
  \citenamefont {Marchini}, \citenamefont {Wintterlin}, \citenamefont
  {Moritz},\ and\ \citenamefont {Greber}}]{MWB08}%
  \BibitemOpen
  \bibfield  {author} {\bibinfo {author} {\bibfnamefont {D.}~\bibnamefont
  {Martoccia}}, \bibinfo {author} {\bibfnamefont {P.~R.}\ \bibnamefont
  {Willmott}}, \bibinfo {author} {\bibfnamefont {T.}~\bibnamefont {Brugger}},
  \bibinfo {author} {\bibfnamefont {M.}~\bibnamefont {Bj\"orck}}, \bibinfo
  {author} {\bibfnamefont {S.}~\bibnamefont {G\"unther}}, \bibinfo {author}
  {\bibfnamefont {C.~M.}\ \bibnamefont {Schlep\"utz}}, \bibinfo {author}
  {\bibfnamefont {A.}~\bibnamefont {Cervellino}}, \bibinfo {author}
  {\bibfnamefont {S.~A.}\ \bibnamefont {Pauli}}, \bibinfo {author}
  {\bibfnamefont {B.~D.}\ \bibnamefont {Patterson}}, \bibinfo {author}
  {\bibfnamefont {S.}~\bibnamefont {Marchini}}, \bibinfo {author}
  {\bibfnamefont {J.}~\bibnamefont {Wintterlin}}, \bibinfo {author}
  {\bibfnamefont {W.}~\bibnamefont {Moritz}}, \ and\ \bibinfo {author}
  {\bibfnamefont {T.}~\bibnamefont {Greber}},\ }\href {\doibase
  10.1103/PhysRevLett.101.126102} {\bibfield  {journal} {\bibinfo  {journal}
  {Phys. Rev. Lett.}\ }\textbf {\bibinfo {volume} {101}},\ \bibinfo {pages}
  {126102} (\bibinfo {year} {2008})}\BibitemShut {NoStop}%
\bibitem [{\citenamefont {V\'azquez~de Parga}\ \emph
  {et~al.}(2008)\citenamefont {V\'azquez~de Parga}, \citenamefont {Calleja},
  \citenamefont {Borca}, \citenamefont {Passeggi}, \citenamefont {Hinarejos},
  \citenamefont {Guinea},\ and\ \citenamefont {Miranda}}]{VCB08}%
  \BibitemOpen
  \bibfield  {author} {\bibinfo {author} {\bibfnamefont {A.~L.}\ \bibnamefont
  {V\'azquez~de Parga}}, \bibinfo {author} {\bibfnamefont {F.}~\bibnamefont
  {Calleja}}, \bibinfo {author} {\bibfnamefont {B.}~\bibnamefont {Borca}},
  \bibinfo {author} {\bibfnamefont {M.~C.~G.}\ \bibnamefont {Passeggi}},
  \bibinfo {author} {\bibfnamefont {J.~J.}\ \bibnamefont {Hinarejos}}, \bibinfo
  {author} {\bibfnamefont {F.}~\bibnamefont {Guinea}}, \ and\ \bibinfo {author}
  {\bibfnamefont {R.}~\bibnamefont {Miranda}},\ }\href {\doibase
  10.1103/PhysRevLett.100.056807} {\bibfield  {journal} {\bibinfo  {journal}
  {Phys. Rev. Lett.}\ }\textbf {\bibinfo {volume} {100}},\ \bibinfo {pages}
  {056807} (\bibinfo {year} {2008})}\BibitemShut {NoStop}%
\bibitem [{\citenamefont {Pan}\ \emph {et~al.}(2009)\citenamefont {Pan},
  \citenamefont {Zhang}, \citenamefont {Shi}, \citenamefont {Sun},
  \citenamefont {Du}, \citenamefont {Liu},\ and\ \citenamefont {Gao}}]{PZS09}%
  \BibitemOpen
  \bibfield  {author} {\bibinfo {author} {\bibfnamefont {Y.}~\bibnamefont
  {Pan}}, \bibinfo {author} {\bibfnamefont {H.}~\bibnamefont {Zhang}}, \bibinfo
  {author} {\bibfnamefont {D.}~\bibnamefont {Shi}}, \bibinfo {author}
  {\bibfnamefont {J.}~\bibnamefont {Sun}}, \bibinfo {author} {\bibfnamefont
  {S.}~\bibnamefont {Du}}, \bibinfo {author} {\bibfnamefont {F.}~\bibnamefont
  {Liu}}, \ and\ \bibinfo {author} {\bibfnamefont {H.-j.}\ \bibnamefont
  {Gao}},\ }\href@noop {} {\bibfield  {journal} {\bibinfo  {journal} {Adv.
  Mater.}\ }\textbf {\bibinfo {volume} {21}},\ \bibinfo {pages} {2777}
  (\bibinfo {year} {2009})}\BibitemShut {NoStop}%
\bibitem [{\citenamefont {Borca}\ \emph {et~al.}(2010)\citenamefont {Borca},
  \citenamefont {Barja}, \citenamefont {Garnica}, \citenamefont {Minniti},
  \citenamefont {Politano}, \citenamefont {Rodriguez-Garc\'ia}, \citenamefont
  {Hinarejos}, \citenamefont {Far\'ias}, \citenamefont {V\'azquez~de Parga},\
  and\ \citenamefont {Miranda}}]{BBG10}%
  \BibitemOpen
  \bibfield  {author} {\bibinfo {author} {\bibfnamefont {B.}~\bibnamefont
  {Borca}}, \bibinfo {author} {\bibfnamefont {S.}~\bibnamefont {Barja}},
  \bibinfo {author} {\bibfnamefont {M.}~\bibnamefont {Garnica}}, \bibinfo
  {author} {\bibfnamefont {M.}~\bibnamefont {Minniti}}, \bibinfo {author}
  {\bibfnamefont {A.}~\bibnamefont {Politano}}, \bibinfo {author}
  {\bibfnamefont {J.~M.}\ \bibnamefont {Rodriguez-Garc\'ia}}, \bibinfo {author}
  {\bibfnamefont {J.~J.}\ \bibnamefont {Hinarejos}}, \bibinfo {author}
  {\bibfnamefont {D.}~\bibnamefont {Far\'ias}}, \bibinfo {author}
  {\bibfnamefont {A.~L.}\ \bibnamefont {V\'azquez~de Parga}}, \ and\ \bibinfo
  {author} {\bibfnamefont {R.}~\bibnamefont {Miranda}},\ }\href@noop {}
  {\bibfield  {journal} {\bibinfo  {journal} {New J. Phys.}\ }\textbf {\bibinfo
  {volume} {12}},\ \bibinfo {pages} {093018} (\bibinfo {year}
  {2010})}\BibitemShut {NoStop}%
\bibitem [{\citenamefont {Martoccia}\ \emph {et~al.}(2010)\citenamefont
  {Martoccia}, \citenamefont {Björck}, \citenamefont {Schlepütz},
  \citenamefont {Brugger}, \citenamefont {Pauli}, \citenamefont {Patterson},
  \citenamefont {Greber},\ and\ \citenamefont {Willmott}}]{MBS10}%
  \BibitemOpen
  \bibfield  {author} {\bibinfo {author} {\bibfnamefont {D.}~\bibnamefont
  {Martoccia}}, \bibinfo {author} {\bibfnamefont {M.}~\bibnamefont {Björck}},
  \bibinfo {author} {\bibfnamefont {C.~M.}\ \bibnamefont {Schlepütz}},
  \bibinfo {author} {\bibfnamefont {T.}~\bibnamefont {Brugger}}, \bibinfo
  {author} {\bibfnamefont {S.~A.}\ \bibnamefont {Pauli}}, \bibinfo {author}
  {\bibfnamefont {B.~D.}\ \bibnamefont {Patterson}}, \bibinfo {author}
  {\bibfnamefont {T.}~\bibnamefont {Greber}}, \ and\ \bibinfo {author}
  {\bibfnamefont {P.~R.}\ \bibnamefont {Willmott}},\ }\href@noop {} {\bibfield
  {journal} {\bibinfo  {journal} {New J. Phys.}\ }\textbf {\bibinfo {volume}
  {12}},\ \bibinfo {pages} {043028} (\bibinfo {year} {2010})}\BibitemShut
  {NoStop}%
\bibitem [{\citenamefont {Moritz}\ \emph {et~al.}(2010)\citenamefont {Moritz},
  \citenamefont {Wang}, \citenamefont {Bocquet}, \citenamefont {Brugger},
  \citenamefont {Greber}, \citenamefont {Wintterlin},\ and\ \citenamefont
  {G\"unther}}]{MWB10}%
  \BibitemOpen
  \bibfield  {author} {\bibinfo {author} {\bibfnamefont {W.}~\bibnamefont
  {Moritz}}, \bibinfo {author} {\bibfnamefont {B.}~\bibnamefont {Wang}},
  \bibinfo {author} {\bibfnamefont {M.-L.}\ \bibnamefont {Bocquet}}, \bibinfo
  {author} {\bibfnamefont {T.}~\bibnamefont {Brugger}}, \bibinfo {author}
  {\bibfnamefont {T.}~\bibnamefont {Greber}}, \bibinfo {author} {\bibfnamefont
  {J.}~\bibnamefont {Wintterlin}}, \ and\ \bibinfo {author} {\bibfnamefont
  {S.}~\bibnamefont {G\"unther}},\ }\href {\doibase
  10.1103/PhysRevLett.104.136102} {\bibfield  {journal} {\bibinfo  {journal}
  {Phys. Rev. Lett.}\ }\textbf {\bibinfo {volume} {104}},\ \bibinfo {pages}
  {136102} (\bibinfo {year} {2010})}\BibitemShut {NoStop}%
\bibitem [{\citenamefont {Politano}\ \emph {et~al.}(2011)\citenamefont
  {Politano}, \citenamefont {Borca}, \citenamefont {Minniti}, \citenamefont
  {Hinarejos}, \citenamefont {V\'azquez~de Parga}, \citenamefont {Far\'ias},\
  and\ \citenamefont {Miranda}}]{PBM11}%
  \BibitemOpen
  \bibfield  {author} {\bibinfo {author} {\bibfnamefont {A.}~\bibnamefont
  {Politano}}, \bibinfo {author} {\bibfnamefont {B.}~\bibnamefont {Borca}},
  \bibinfo {author} {\bibfnamefont {M.}~\bibnamefont {Minniti}}, \bibinfo
  {author} {\bibfnamefont {J.~J.}\ \bibnamefont {Hinarejos}}, \bibinfo {author}
  {\bibfnamefont {A.~L.}\ \bibnamefont {V\'azquez~de Parga}}, \bibinfo {author}
  {\bibfnamefont {D.}~\bibnamefont {Far\'ias}}, \ and\ \bibinfo {author}
  {\bibfnamefont {R.}~\bibnamefont {Miranda}},\ }\href {\doibase
  10.1103/PhysRevB.84.035450} {\bibfield  {journal} {\bibinfo  {journal} {Phys.
  Rev. B}\ }\textbf {\bibinfo {volume} {84}},\ \bibinfo {pages} {035450}
  (\bibinfo {year} {2011})}\BibitemShut {NoStop}%
\bibitem [{\citenamefont {Günther}\ \emph {et~al.}(2011)\citenamefont
  {Günther}, \citenamefont {Dänhardt}, \citenamefont {Wang}, \citenamefont
  {Bocquet}, \citenamefont {Schmitt},\ and\ \citenamefont
  {Wintterlin}}]{GDW11}%
  \BibitemOpen
  \bibfield  {author} {\bibinfo {author} {\bibfnamefont {S.}~\bibnamefont
  {Günther}}, \bibinfo {author} {\bibfnamefont {S.}~\bibnamefont
  {Dänhardt}}, \bibinfo {author} {\bibfnamefont {B.}~\bibnamefont {Wang}},
  \bibinfo {author} {\bibfnamefont {M.-L.}\ \bibnamefont {Bocquet}}, \bibinfo
  {author} {\bibfnamefont {S.}~\bibnamefont {Schmitt}}, \ and\ \bibinfo
  {author} {\bibfnamefont {J.}~\bibnamefont {Wintterlin}},\ }\href@noop {}
  {\bibfield  {journal} {\bibinfo  {journal} {Nanolett.}\ }\textbf {\bibinfo
  {volume} {11}},\ \bibinfo {pages} {1895} (\bibinfo {year}
  {2011})}\BibitemShut {NoStop}%
\bibitem [{\citenamefont {Armbrust}\ \emph {et~al.}(2012)\citenamefont
  {Armbrust}, \citenamefont {G\"udde}, \citenamefont {Jakob},\ and\
  \citenamefont {H\"ofer}}]{AGJ12}%
  \BibitemOpen
  \bibfield  {author} {\bibinfo {author} {\bibfnamefont {N.}~\bibnamefont
  {Armbrust}}, \bibinfo {author} {\bibfnamefont {J.}~\bibnamefont {G\"udde}},
  \bibinfo {author} {\bibfnamefont {P.}~\bibnamefont {Jakob}}, \ and\ \bibinfo
  {author} {\bibfnamefont {U.}~\bibnamefont {H\"ofer}},\ }\href {\doibase
  10.1103/PhysRevLett.108.056801} {\bibfield  {journal} {\bibinfo  {journal}
  {Phys. Rev. Lett.}\ }\textbf {\bibinfo {volume} {108}},\ \bibinfo {pages}
  {056801} (\bibinfo {year} {2012})}\BibitemShut {NoStop}%
\bibitem [{\citenamefont {Politano}\ \emph {et~al.}(2013)\citenamefont
  {Politano}, \citenamefont {Campi}, \citenamefont {Formoso},\ and\
  \citenamefont {Chiarello}}]{PCF13}%
  \BibitemOpen
  \bibfield  {author} {\bibinfo {author} {\bibfnamefont {A.}~\bibnamefont
  {Politano}}, \bibinfo {author} {\bibfnamefont {D.}~\bibnamefont {Campi}},
  \bibinfo {author} {\bibfnamefont {V.}~\bibnamefont {Formoso}}, \ and\
  \bibinfo {author} {\bibfnamefont {G.}~\bibnamefont {Chiarello}},\ }\href@noop
  {} {\bibfield  {journal} {\bibinfo  {journal} {Phys. Chem. Chem. Phys.}\
  }\textbf {\bibinfo {volume} {15}},\ \bibinfo {pages} {11356} (\bibinfo {year}
  {2013})}\BibitemShut {NoStop}%
\bibitem [{\citenamefont {Loginova}\ \emph {et~al.}(2009)\citenamefont
  {Loginova}, \citenamefont {Nie}, \citenamefont {Th{\"u}rmer}, \citenamefont
  {Bartelt},\ and\ \citenamefont {McCarty}}]{LNT09}%
  \BibitemOpen
  \bibfield  {author} {\bibinfo {author} {\bibfnamefont {E.}~\bibnamefont
  {Loginova}}, \bibinfo {author} {\bibfnamefont {S.}~\bibnamefont {Nie}},
  \bibinfo {author} {\bibfnamefont {K.}~\bibnamefont {Th{\"u}rmer}}, \bibinfo
  {author} {\bibfnamefont {N.~C.}\ \bibnamefont {Bartelt}}, \ and\ \bibinfo
  {author} {\bibfnamefont {K.~F.}\ \bibnamefont {McCarty}},\ }\href@noop {}
  {\bibfield  {journal} {\bibinfo  {journal} {Phys. Rev. B}\ }\textbf {\bibinfo
  {volume} {80}},\ \bibinfo {pages} {085430} (\bibinfo {year}
  {2009})}\BibitemShut {NoStop}%
\bibitem [{\citenamefont {Cazzanelli}\ \emph {et~al.}(2013)\citenamefont
  {Cazzanelli}, \citenamefont {Caruso}, \citenamefont {Castriota},
  \citenamefont {Marino}, \citenamefont {Politano}, \citenamefont {Chiarello},
  \citenamefont {Giarola},\ and\ \citenamefont {Mariotto}}]{CCC13}%
  \BibitemOpen
  \bibfield  {author} {\bibinfo {author} {\bibfnamefont {E.}~\bibnamefont
  {Cazzanelli}}, \bibinfo {author} {\bibfnamefont {T.}~\bibnamefont {Caruso}},
  \bibinfo {author} {\bibfnamefont {M.}~\bibnamefont {Castriota}}, \bibinfo
  {author} {\bibfnamefont {A.}~\bibnamefont {Marino}}, \bibinfo {author}
  {\bibfnamefont {A.}~\bibnamefont {Politano}}, \bibinfo {author}
  {\bibfnamefont {G.}~\bibnamefont {Chiarello}}, \bibinfo {author}
  {\bibfnamefont {M.}~\bibnamefont {Giarola}}, \ and\ \bibinfo {author}
  {\bibfnamefont {G.}~\bibnamefont {Mariotto}},\ }\href@noop {} {\bibfield
  {journal} {\bibinfo  {journal} {J. Raman Spectrosc.}\ }\textbf {\bibinfo
  {volume} {44}},\ \bibinfo {pages} {1393} (\bibinfo {year}
  {2013})}\BibitemShut {NoStop}%
\bibitem [{Note2()}]{Note2}%
  \BibitemOpen
  \bibinfo {note} {The phonon dispersion along $M K$ was recorded with $E_I =$
  32 eV. However, the cross-section of phonon modes in this scattering
  conditions became so weak that a comparison with data acquired by using
  $E_I=$ 20 eV would not be meaningful.}\BibitemShut {Stop}%
\bibitem [{\citenamefont {Falkovsky}(2008)}]{F08}%
  \BibitemOpen
  \bibfield  {author} {\bibinfo {author} {\bibfnamefont {L.}~\bibnamefont
  {Falkovsky}},\ }\href@noop {} {\bibfield  {journal} {\bibinfo  {journal}
  {Phys. Lett. A}\ }\textbf {\bibinfo {volume} {372}},\ \bibinfo {pages} {5189}
  (\bibinfo {year} {2008})}\BibitemShut {NoStop}%
\bibitem [{\citenamefont {Viola~Kusminskiy}\ \emph {et~al.}(2009)\citenamefont
  {Viola~Kusminskiy}, \citenamefont {Campbell},\ and\ \citenamefont
  {Castro~Neto}}]{VCC09}%
  \BibitemOpen
  \bibfield  {author} {\bibinfo {author} {\bibfnamefont {S.}~\bibnamefont
  {Viola~Kusminskiy}}, \bibinfo {author} {\bibfnamefont {D.~K.}\ \bibnamefont
  {Campbell}}, \ and\ \bibinfo {author} {\bibfnamefont {A.~H.}\ \bibnamefont
  {Castro~Neto}},\ }\href {\doibase 10.1103/PhysRevB.80.035401} {\bibfield
  {journal} {\bibinfo  {journal} {Phys. Rev. B}\ }\textbf {\bibinfo {volume}
  {80}},\ \bibinfo {pages} {035401} (\bibinfo {year} {2009})}\BibitemShut
  {NoStop}%
\bibitem [{\citenamefont {Wirtz}\ and\ \citenamefont {Rubio}(2004)}]{WR04}%
  \BibitemOpen
  \bibfield  {author} {\bibinfo {author} {\bibfnamefont {L.}~\bibnamefont
  {Wirtz}}\ and\ \bibinfo {author} {\bibfnamefont {A.}~\bibnamefont {Rubio}},\
  }\href@noop {} {\bibfield  {journal} {\bibinfo  {journal} {Solid State
  Commun.}\ }\textbf {\bibinfo {volume} {131}},\ \bibinfo {pages} {141}
  (\bibinfo {year} {2004})}\BibitemShut {NoStop}%
\bibitem [{\citenamefont {Lazzeri}\ \emph {et~al.}(2008)\citenamefont
  {Lazzeri}, \citenamefont {Attaccalite}, \citenamefont {Wirtz},\ and\
  \citenamefont {Mauri}}]{LAW08}%
  \BibitemOpen
  \bibfield  {author} {\bibinfo {author} {\bibfnamefont {M.}~\bibnamefont
  {Lazzeri}}, \bibinfo {author} {\bibfnamefont {C.}~\bibnamefont
  {Attaccalite}}, \bibinfo {author} {\bibfnamefont {L.}~\bibnamefont {Wirtz}},
  \ and\ \bibinfo {author} {\bibfnamefont {F.}~\bibnamefont {Mauri}},\ }\href
  {\doibase 10.1103/PhysRevB.78.081406} {\bibfield  {journal} {\bibinfo
  {journal} {Phys. Rev. B}\ }\textbf {\bibinfo {volume} {78}},\ \bibinfo
  {pages} {081406} (\bibinfo {year} {2008})}\BibitemShut {NoStop}%
\bibitem [{\citenamefont {Allard}\ and\ \citenamefont {Wirtz}(2010)}]{AW10}%
  \BibitemOpen
  \bibfield  {author} {\bibinfo {author} {\bibfnamefont {A.}~\bibnamefont
  {Allard}}\ and\ \bibinfo {author} {\bibfnamefont {L.}~\bibnamefont {Wirtz}},\
  }\href@noop {} {\bibfield  {journal} {\bibinfo  {journal} {Nano lett.}\
  }\textbf {\bibinfo {volume} {10}},\ \bibinfo {pages} {4335} (\bibinfo {year}
  {2010})}\BibitemShut {NoStop}%
\bibitem [{\citenamefont {Mohiuddin}\ \emph {et~al.}(2009)\citenamefont
  {Mohiuddin}, \citenamefont {Lombardo}, \citenamefont {Nair}, \citenamefont
  {Bonetti}, \citenamefont {Savini}, \citenamefont {Jalil}, \citenamefont
  {Bonini}, \citenamefont {Basko}, \citenamefont {Galiotis}, \citenamefont
  {Marzari}, \citenamefont {Novoselov}, \citenamefont {Geim},\ and\
  \citenamefont {Ferrari}}]{MLN09}%
  \BibitemOpen
  \bibfield  {author} {\bibinfo {author} {\bibfnamefont {T.~M.~G.}\
  \bibnamefont {Mohiuddin}}, \bibinfo {author} {\bibfnamefont {A.}~\bibnamefont
  {Lombardo}}, \bibinfo {author} {\bibfnamefont {R.~R.}\ \bibnamefont {Nair}},
  \bibinfo {author} {\bibfnamefont {A.}~\bibnamefont {Bonetti}}, \bibinfo
  {author} {\bibfnamefont {G.}~\bibnamefont {Savini}}, \bibinfo {author}
  {\bibfnamefont {R.}~\bibnamefont {Jalil}}, \bibinfo {author} {\bibfnamefont
  {N.}~\bibnamefont {Bonini}}, \bibinfo {author} {\bibfnamefont {D.~M.}\
  \bibnamefont {Basko}}, \bibinfo {author} {\bibfnamefont {C.}~\bibnamefont
  {Galiotis}}, \bibinfo {author} {\bibfnamefont {N.}~\bibnamefont {Marzari}},
  \bibinfo {author} {\bibfnamefont {K.~S.}\ \bibnamefont {Novoselov}}, \bibinfo
  {author} {\bibfnamefont {A.~K.}\ \bibnamefont {Geim}}, \ and\ \bibinfo
  {author} {\bibfnamefont {A.~C.}\ \bibnamefont {Ferrari}},\ }\href {\doibase
  10.1103/PhysRevB.79.205433} {\bibfield  {journal} {\bibinfo  {journal} {Phys.
  Rev. B}\ }\textbf {\bibinfo {volume} {79}},\ \bibinfo {pages} {205433}
  (\bibinfo {year} {2009})}\BibitemShut {NoStop}%
\bibitem [{\citenamefont {Shikin}\ \emph {et~al.}(1998)\citenamefont {Shikin},
  \citenamefont {Farias},\ and\ \citenamefont {Rieder}}]{SFR98}%
  \BibitemOpen
  \bibfield  {author} {\bibinfo {author} {\bibfnamefont {A.}~\bibnamefont
  {Shikin}}, \bibinfo {author} {\bibfnamefont {D.}~\bibnamefont {Farias}}, \
  and\ \bibinfo {author} {\bibfnamefont {K.}~\bibnamefont {Rieder}},\
  }\href@noop {} {\bibfield  {journal} {\bibinfo  {journal} {EPL}\ }\textbf
  {\bibinfo {volume} {44}},\ \bibinfo {pages} {44} (\bibinfo {year}
  {1998})}\BibitemShut {NoStop}%
\bibitem [{\citenamefont {Rokuta}\ \emph {et~al.}(1997)\citenamefont {Rokuta},
  \citenamefont {Hasegawa}, \citenamefont {Suzuki}, \citenamefont {Gamou},
  \citenamefont {Oshima},\ and\ \citenamefont {Nagashima}}]{RHS97}%
  \BibitemOpen
  \bibfield  {author} {\bibinfo {author} {\bibfnamefont {E.}~\bibnamefont
  {Rokuta}}, \bibinfo {author} {\bibfnamefont {Y.}~\bibnamefont {Hasegawa}},
  \bibinfo {author} {\bibfnamefont {K.}~\bibnamefont {Suzuki}}, \bibinfo
  {author} {\bibfnamefont {Y.}~\bibnamefont {Gamou}}, \bibinfo {author}
  {\bibfnamefont {C.}~\bibnamefont {Oshima}}, \ and\ \bibinfo {author}
  {\bibfnamefont {A.}~\bibnamefont {Nagashima}},\ }\href {\doibase
  10.1103/PhysRevLett.79.4609} {\bibfield  {journal} {\bibinfo  {journal}
  {Phys. Rev. Lett.}\ }\textbf {\bibinfo {volume} {79}},\ \bibinfo {pages}
  {4609} (\bibinfo {year} {1997})}\BibitemShut {NoStop}%
\bibitem [{\citenamefont {Vogt}\ \emph {et~al.}(2012)\citenamefont {Vogt},
  \citenamefont {De~Padova}, \citenamefont {Quaresima}, \citenamefont {Avila},
  \citenamefont {Frantzeskakis}, \citenamefont {Asensio}, \citenamefont
  {Resta}, \citenamefont {Ealet},\ and\ \citenamefont {Le~Lay}}]{VDQ12}%
  \BibitemOpen
  \bibfield  {author} {\bibinfo {author} {\bibfnamefont {P.}~\bibnamefont
  {Vogt}}, \bibinfo {author} {\bibfnamefont {P.}~\bibnamefont {De~Padova}},
  \bibinfo {author} {\bibfnamefont {C.}~\bibnamefont {Quaresima}}, \bibinfo
  {author} {\bibfnamefont {J.}~\bibnamefont {Avila}}, \bibinfo {author}
  {\bibfnamefont {E.}~\bibnamefont {Frantzeskakis}}, \bibinfo {author}
  {\bibfnamefont {M.~C.}\ \bibnamefont {Asensio}}, \bibinfo {author}
  {\bibfnamefont {A.}~\bibnamefont {Resta}}, \bibinfo {author} {\bibfnamefont
  {B.}~\bibnamefont {Ealet}}, \ and\ \bibinfo {author} {\bibfnamefont
  {G.}~\bibnamefont {Le~Lay}},\ }\href {\doibase
  10.1103/PhysRevLett.108.155501} {\bibfield  {journal} {\bibinfo  {journal}
  {Phys. Rev. Lett.}\ }\textbf {\bibinfo {volume} {108}},\ \bibinfo {pages}
  {155501} (\bibinfo {year} {2012})}\BibitemShut {NoStop}%
\bibitem [{\citenamefont {Fleurence}\ \emph {et~al.}(2012)\citenamefont
  {Fleurence}, \citenamefont {Friedlein}, \citenamefont {Ozaki}, \citenamefont
  {Kawai}, \citenamefont {Wang},\ and\ \citenamefont
  {Yamada-Takamura}}]{FFR12}%
  \BibitemOpen
  \bibfield  {author} {\bibinfo {author} {\bibfnamefont {A.}~\bibnamefont
  {Fleurence}}, \bibinfo {author} {\bibfnamefont {R.}~\bibnamefont
  {Friedlein}}, \bibinfo {author} {\bibfnamefont {T.}~\bibnamefont {Ozaki}},
  \bibinfo {author} {\bibfnamefont {H.}~\bibnamefont {Kawai}}, \bibinfo
  {author} {\bibfnamefont {Y.}~\bibnamefont {Wang}}, \ and\ \bibinfo {author}
  {\bibfnamefont {Y.}~\bibnamefont {Yamada-Takamura}},\ }\href {\doibase
  10.1103/PhysRevLett.108.245501} {\bibfield  {journal} {\bibinfo  {journal}
  {Phys. Rev. Lett.}\ }\textbf {\bibinfo {volume} {108}},\ \bibinfo {pages}
  {245501} (\bibinfo {year} {2012})}\BibitemShut {NoStop}%
\bibitem [{\citenamefont {Wilkes}\ \emph {et~al.}(1987)\citenamefont {Wilkes},
  \citenamefont {Palmer},\ and\ \citenamefont {Willis}}]{WPW87}%
  \BibitemOpen
  \bibfield  {author} {\bibinfo {author} {\bibfnamefont {J.}~\bibnamefont
  {Wilkes}}, \bibinfo {author} {\bibfnamefont {R.}~\bibnamefont {Palmer}}, \
  and\ \bibinfo {author} {\bibfnamefont {R.}~\bibnamefont {Willis}},\
  }\href@noop {} {\bibfield  {journal} {\bibinfo  {journal} {J. Electron.
  Spectrosc. Relat. Phenom.}\ }\textbf {\bibinfo {volume} {44}},\ \bibinfo
  {pages} {355} (\bibinfo {year} {1987})}\BibitemShut {NoStop}%
\bibitem [{\citenamefont {Mohr}\ \emph {et~al.}(2007)\citenamefont {Mohr},
  \citenamefont {Maultzsch}, \citenamefont {Dobard\ifmmode \check{z}\else
  \v{z}\fi{}i\ifmmode~\acute{c}\else \'{c}\fi{}}, \citenamefont {Reich},
  \citenamefont {Milo\ifmmode \check{s}\else
  \v{s}\fi{}evi\ifmmode~\acute{c}\else \'{c}\fi{}}, \citenamefont
  {Damnjanovi\ifmmode~\acute{c}\else \'{c}\fi{}}, \citenamefont {Bosak},
  \citenamefont {Krisch},\ and\ \citenamefont {Thomsen}}]{MMD07}%
  \BibitemOpen
  \bibfield  {author} {\bibinfo {author} {\bibfnamefont {M.}~\bibnamefont
  {Mohr}}, \bibinfo {author} {\bibfnamefont {J.}~\bibnamefont {Maultzsch}},
  \bibinfo {author} {\bibfnamefont {E.}~\bibnamefont {Dobard\ifmmode
  \check{z}\else \v{z}\fi{}i\ifmmode~\acute{c}\else \'{c}\fi{}}}, \bibinfo
  {author} {\bibfnamefont {S.}~\bibnamefont {Reich}}, \bibinfo {author}
  {\bibfnamefont {I.}~\bibnamefont {Milo\ifmmode \check{s}\else
  \v{s}\fi{}evi\ifmmode~\acute{c}\else \'{c}\fi{}}}, \bibinfo {author}
  {\bibfnamefont {M.}~\bibnamefont {Damnjanovi\ifmmode~\acute{c}\else
  \'{c}\fi{}}}, \bibinfo {author} {\bibfnamefont {A.}~\bibnamefont {Bosak}},
  \bibinfo {author} {\bibfnamefont {M.}~\bibnamefont {Krisch}}, \ and\ \bibinfo
  {author} {\bibfnamefont {C.}~\bibnamefont {Thomsen}},\ }\href {\doibase
  10.1103/PhysRevB.76.035439} {\bibfield  {journal} {\bibinfo  {journal} {Phys.
  Rev. B}\ }\textbf {\bibinfo {volume} {76}},\ \bibinfo {pages} {035439}
  (\bibinfo {year} {2007})}\BibitemShut {NoStop}%
\bibitem [{\citenamefont {Rocca}(1995)}]{R95}%
  \BibitemOpen
  \bibfield  {author} {\bibinfo {author} {\bibfnamefont {M.}~\bibnamefont
  {Rocca}},\ }\href {\doibase 10.1016/0167-5729(95)00004-6} {\bibfield
  {journal} {\bibinfo  {journal} {Surf. Sci. Rep.}\ }\textbf {\bibinfo {volume}
  {22}},\ \bibinfo {pages} {1 } (\bibinfo {year} {1995})}\BibitemShut {NoStop}%
\bibitem [{\citenamefont {Politano}\ \emph {et~al.}(2009)\citenamefont
  {Politano}, \citenamefont {Formoso}, \citenamefont {Colavita},\ and\
  \citenamefont {Chiarello}}]{PFC09}%
  \BibitemOpen
  \bibfield  {author} {\bibinfo {author} {\bibfnamefont {A.}~\bibnamefont
  {Politano}}, \bibinfo {author} {\bibfnamefont {V.}~\bibnamefont {Formoso}},
  \bibinfo {author} {\bibfnamefont {E.}~\bibnamefont {Colavita}}, \ and\
  \bibinfo {author} {\bibfnamefont {G.}~\bibnamefont {Chiarello}},\ }\href
  {\doibase 10.1103/PhysRevB.79.045426} {\bibfield  {journal} {\bibinfo
  {journal} {Phys. Rev. B}\ }\textbf {\bibinfo {volume} {79}},\ \bibinfo
  {pages} {045426} (\bibinfo {year} {2009})}\BibitemShut {NoStop}%
\bibitem [{\citenamefont {Politano}\ \emph {et~al.}(2008)\citenamefont
  {Politano}, \citenamefont {Formoso},\ and\ \citenamefont
  {Chiarello}}]{PFC08}%
  \BibitemOpen
  \bibfield  {author} {\bibinfo {author} {\bibfnamefont {A.}~\bibnamefont
  {Politano}}, \bibinfo {author} {\bibfnamefont {V.}~\bibnamefont {Formoso}}, \
  and\ \bibinfo {author} {\bibfnamefont {G.}~\bibnamefont {Chiarello}},\
  }\href@noop {} {\bibfield  {journal} {\bibinfo  {journal} {Plasmonics}\
  }\textbf {\bibinfo {volume} {3}},\ \bibinfo {pages} {165} (\bibinfo {year}
  {2008})}\BibitemShut {NoStop}%
\bibitem [{\citenamefont {Huang}\ \emph {et~al.}(2009)\citenamefont {Huang},
  \citenamefont {Yan}, \citenamefont {Chen}, \citenamefont {Song},
  \citenamefont {Heinz},\ and\ \citenamefont {Hone}}]{MHC09}%
  \BibitemOpen
  \bibfield  {author} {\bibinfo {author} {\bibfnamefont {M.}~\bibnamefont
  {Huang}}, \bibinfo {author} {\bibfnamefont {H.}~\bibnamefont {Yan}}, \bibinfo
  {author} {\bibfnamefont {C.}~\bibnamefont {Chen}}, \bibinfo {author}
  {\bibfnamefont {D.}~\bibnamefont {Song}}, \bibinfo {author} {\bibfnamefont
  {T.~F.}\ \bibnamefont {Heinz}}, \ and\ \bibinfo {author} {\bibfnamefont
  {J.}~\bibnamefont {Hone}},\ }\href@noop {} {\bibfield  {journal} {\bibinfo
  {journal} {Proc. Natl. Acad. Sci.}\ }\textbf {\bibinfo {volume} {106}},\
  \bibinfo {pages} {7304} (\bibinfo {year} {2009})}\BibitemShut {NoStop}%
\end{thebibliography}%

%\newpage
%\pagebreak
%\onecolumngrid
%\vspace{0.2in}
%\begin{center}
%{\bf \large  Symmetries and selection rules in experimental measurements of the phonon
%spectrum of graphene and related materials -
%Supplementary materials - }
%\end{center}
%\vspace{0.1in}

%\renewcommand{\thetable}{S\Roman{table}}
%\renewcommand{\thefigure}{S\arabic{figure}}
%\renewcommand{\thesubsection}{S\arabic{subsection}}
%\renewcommand{\theequation}{S\arabic{equation}}

%\setcounter{secnumdepth}{1}
%\setcounter{equation}{0}
%\setcounter{figure}{0}
%\setcounter{section}{0}

\end{document}